\documentclass[prc,twocolumn,nofootinbib,superscriptaddress,showpacs]{revtex4}
\usepackage{graphicx,amsmath,amssymb,bm,multirow}
\usepackage{color}
\usepackage{amscd}

\newcommand{\be}[1]{\begin{equation}\label{#1}}
\newcommand{\ee}{\end{equation}}
\newcommand{\vlowk}{V_{{\rm low}\,k}}

\newcommand{\fm}{\, \text{fm}}
\newcommand{\fmi}{\, \text{fm}^{-1}}
\newcommand{\fmicube}{\, \text{fm}^{-3}}
\newcommand{\mev}{\, \text{MeV}}
\newcommand{\kev}{\, \text{keV}}
\newcommand{\gcubecm}{\text{g} \cdot \text{cm}^{-3}}

\newcommand{\elemA}[2]{\ensuremath{{}^{#1}}\textrm{#2}}

\newcommand{\partialwave}[3]{\ensuremath{{}^{#1}}\textrm{#2}_{#3}}

\begin{document}

\title{On the superfluid properties of the inner crust of neutron stars}

\author{Alessandro Pastore}
\email{pastore@inpl.in2p3.fr}
\affiliation{Department of Physics, Post Office Box 35 (YFL), 
FI-40014 University of Jyv\"askyl\"a, Finland}
\affiliation{Universit\'e de Lyon, F-69003 Lyon, France; Universit\'e Lyon 1, 
43 Bd. du 11 Novembre 1918, F-69622 Villeurbanne
cedex, France CNRS-IN2P3, UMR 5822, Institut de Physique Nucl\'{e}aire de Lyon}
\author{Simone Baroni}
\email{baroni@triumf.ca}
\affiliation{TRIUMF, 4004 Wesbrook Mall, Vancouver BC, V6T 2A3, Canada}
\author{Cristina Losa}
\email{losacris@sissa.it}
\affiliation{International School for Advanced Studies, SISSA Via Bonomea 265 34136 Trieste, Italy}

\date{\today}

\begin{abstract}
We investigated the superfluid properties of the inner crust of neutron stars,
solving the Hartree-Fock-Bogoliubov equations in spherical Wigner-Seitz cells.
Using realistic two-body interactions in the pairing channel, we studied in detail
the Cooper-pair and the pairing-field spatial properties, together with the effect of the 
proton clusters on the neutron pairing gap.
Calculations with effective pairing interactions are also presented, showing significant
discrepancies with the results obtained with realistic pairing forces.
At variance with recent studies on finite nuclei, the neutron coherence length is found
to depend on the strength of the pairing interaction, even inside the nucleus.
We also show that the Wigner-Seitz approximation breaks down in the innermost regions of the 
inner crust, already at baryonic densities $\rho_b \ge 8 \cdot 10^{+13}$ g $\cdot$ cm$^{-3}$.
\end{abstract}


\maketitle

\section{\label{sec:intro}Introduction}

The inner crust of neutron stars \cite{Chamel_2008} 
offers a unique opportunity to test nuclear-structure models.
Its extremely neutron-rich environment represents a very strong challenge to theories that have 
been developed for finite nuclei.
In particular, the Energy-Density Functional (EDF) method is able to describe
to good accuracy properties of medium-mass to heavy nuclei \cite{Bender_2003, Bertsch_2009}
and it has been applied to the inner crust of neutron stars since the seminal 
work of Negele and Vautherin \cite{Negele_1973}.
They divided the inner crust of neutron stars into independent spherical
Wigner-Seitz (WS) cells \cite{WS_1933},
each of them representing a inner-crust region of a given density.

To deal with a large number of nucleons, the EDF method relies on effective interactions that are fitted 
also to stable-nuclei experimental data.
Recently, independent studies have been carried out by different groups 
\cite{Lesinski_2009,Baroni_2010,Duguet_2010,Lesinski_2011} to improve
the connection of EDF theories to basic nuclear forces. Particular attention has been paid 
to the pairing correlations that are responsible for the superfluid properties of the nucleus.
These studies use, in the pairing channel, phase-shift-equivalent interactions 
(so-called realistic interactions) evolved to low momentum through 
Renormalization-Group (RG) techniques 
\cite{Bogner_2003, Bogner_2007, Bogner_2007b, Bogner_2010}.

The RG evolution is helpful in many respects. Not only it softens the original hard-core
interactions, making the calculations feasible, but also it helps increase the 
EDF calculation reliability.
This is because the hard-core potentials connect high-energy states, 
whose effective mass is not well described
by the phenomenological EDF functionals, leading to reliability issues \cite{Hebeler_2009}. 
Low-momentum interactions do not probe states too much high in energy.

Along these lines, in this work we study the superfluid properties of the inner crust of neutron stars, 
adopting the WS approximation and solving the Hartree-Fock-Bogoliubov (HFB) equations with low-momentum 
realistic interactions ($\vlowk$) in the pairing channel.
The pairing interaction matrix elements are computed at first order only, 
leaving higher-order correlations for future investigations.
For comparison, we also perform calculations with effective pairing interactions, 
namely the Gogny D1 interaction \cite{Gogny_1980} 
and a Density-Dependent Delta Interaction (DDDI) \cite{Bertsch_1991},
which have both been used in the past as pairing interactions in WS calculations 
\cite{Than_2011, Pizzochero_2002, Grill_2011, Baldo_2005, Fortin_2011}.
The superfluid properties obtained with these effective pairing forces turn out to differ
substancially from those obtained with realistic pairing potentials
(see Sects.~\ref{Sect:results_pairing_gaps} and \ref{Sect:results_pairing_field}).

A comparison with the full band theory \cite{Chamel_2007,Chamel_2010} has shown that 
the WS approximation can reproduce well ground-state properties of the outermost regions 
of the inner crust. Its validity in the regions
closer to the star core, where the clusters nearly touch each other, is still under debate. In this
work we find that the WS approximation starts to break down at baryonic densities 
$\rho_b \approx 8 \cdot 10^{13}$ g $\cdot$ cm$^{-3}$, 
where the protons leak out of the center of the cell and
boundary-condition effects start kicking in. 
See Sect.~\ref{Subsect:innercrust} for a detailed discussion (see also Ref.~\cite{Baldo_2006}).

We study 11 different baryonic-density regions of the inner crust, from 
$\rho_b \approx 5 \cdot 10^{11}$ g $\cdot$ cm$^{-3} \approx 0.0018 \rho_0$ 
to $\rho_b \approx 10^{14}$ $\gcubecm \approx 0.35 \rho_0$, 
with the saturation density $\rho_0 = 2.8 \cdot 10^{14}$ $\gcubecm = 0.16 \fmicube$.
The corresponding WS-cell properties are shown in Table~(\ref{tabWS}) and have been 
taken from previous Hartree-Fock (HF) energy-minimization calculations \cite{Negele_1973}.
Recently, Hartree-Fock-Bardeen-Cooper-Schrieffer (HFBCS) 
\cite{Baldo_2005, Baldo_2006} and HFB \cite{Grill_2011} minimization procedures have been carried out,
obtaining ($R_{WS}, Z$) configurations that differ substancially
from those in Ref.~\cite{Negele_1973}. The actual cluster configurations in the inner crust
still represent an open question and the energy-minimization calculations are very sensitive to the
functionals used. 
However, as we show in this work, the superfluid properties of the system
turn out to be rather independent of the ($R_{WS}, Z$) configurations adopted for a given density region.

In Sect.~\ref{Sect:calc_details} we present the details of the calculations, followed
by the results in Sect.~\ref{Sect:results}.
Sect.~\ref{Sect:results_density} deals with the density profiles and in particular with the instabilities
of the proton density for regions close to the star core. The effect of the proton clusters on the 
neutron pairing gap is discussed in Sect.~\ref{Sect:results_pairing_gaps}, where results for
Infinite Neutron Matter (INM) are compared to those in the inner crust. The spatial properties of the 
Cooper-pair wave function and of the pairing field are treated in
Sects.~\ref{Sect:results_cooperpair} and \ref{Sect:results_pairing_field}.
Conclusions and outlook are presented in Sect.~\ref{Sect:conclusions}.

\section{Calculation details} \label{Sect:calc_details}

\subsection{Inner crust of neutron stars}\label{Subsect:innercrust}

The self-consistent HFB equations \cite{Ring_1980} 
are solved in each representative WS cell on a spherical mesh. 
The mesh step is $0.2\fm$ for the cells 11 through 8 and $0.1\fm$ for the higher-density
cells (see Table~(\ref{tabWS})).
The single-particle wave functions are expanded on a spherical Bessel basis
with a momentum cutoff $k_{max}=4\fmi$. 
This corresponds to an HFB model-space energy cutoff of about 
$\hbar^2k_{max}^2/2m \approx 320\mev$. 
Our calculations are stable with respect to an increase of the model space
and to a decrease of the mesh size.
The WS-cell approximation relies on the fact that the structure of the inner crust of neutron stars
is recovered by a repetition in space of the WS cell. This requires
the neutron density at the edge of the cell to be finite and to match that of
the neighbor cells. This can be achieved by imposing the following 
Dirichlet-Neumann mixed boundary conditions
\cite{Negele_1973}: (i) even-parity wave functions vanish at $R=R_{WS}$; (ii) the first derivative
of odd-parity wave funtions vanishes at $R=R_{WS}$.
We call them Boundary Conditions Even (BCE), in contrast to the Boundary Conditions Odd (BCO)
where the two parity states are treated in the opposite way. 

We use a Skyrme functional to build the single-particle Hamiltonian $h$
and then we let the particles interact pairwise in the pairing channel.
The two-body matrix elements of the pairing interaction in the $J=0,T=1$ channel enter the 
neutron-neutron and proton-proton gap equations, whose solutions provide the matrix elements
of the state-dependent gap matrix $\Delta$.
The latter, in turn, enters the HFB equations

\begin{eqnarray}\label{HFBeq}
  \sum_{n'}(h_{n'nlj}^q- \varepsilon_{F,q})U^{i,q}_{n'lj}+\sum_{n'}\Delta_{nn'lj}^qV^{i,q}_{n'lj}=E^{q}_{ilj}U^{i,q}_{nlj} & & 
     \nonumber \\
  \sum_{n'}\Delta^q_{nn'lj}U^{i,q}_{n'lj} -\sum_{n'}(h^q_{n'nlj}- \varepsilon_{F,q})V^{i,q}_{n'lj}  =E^{q}_{ilj}V^{i,q}_{nlj} & & 
      \nonumber \\
\end{eqnarray}
where $\varepsilon_{F,q}$ is the Fermi energy and $q$ stands for neutrons and protons.
We used the standard notation $nlj$ for the spherical single-particle
states with radial quantum number $n$, orbital angular momentum $l$ and
total angular momentum $j$. 
$U^{i,q}_{nlj}$ and $V^{i,q}_{nlj}$ are the Bogoliubov amplitudes for the $i$-th quasiparticle
of energy $E^{q}_{ilj}$.

When presenting the results for the HFB neutron pairing gaps, we
show the Lowest-quasiparticle-energy Canonical State (LCS) pairing gaps \cite{Lesinski_2009}.
The LCS gap is the diagonal matrix element of the gap matrix for the canonical
state $n_a l_a j_a$ with the lowest canonical quasiparticle energy
\begin{equation}\label{canonicalqpe}
E_{n_a l_a j_a}= \sqrt{(\varepsilon_{n_a l_a j_a} - \varepsilon_{F})^2 + \Delta_{n_a n_a l_a j_a}^2} \, ,
\end{equation}
where $\varepsilon_{n_a l_a j_a}$ is the canonical single-particle energy. We dropped the isospin
index $q$ in Eq.~\ref{canonicalqpe}.

\begin{table}
\begin{center}
\begin{tabular}{ccccccc}
\hline
Zone & Element & Z & N& $R_{WS}$ [fm] & $\rho_b $ [g $\cdot$ cm$^{-3}$] & $k_{F,n}$ [fm$^{-1}$]\\
\hline
11 & $^{180}$Zr & 40 & 140 & 53.6 & $4.67\cdot 10^{11}$ & 0.12\\
10 & $^{200}$Zr & 40 & 160 & 49.2 & $6.69\cdot 10^{11}$ &0.15 \\
 9 & $^{250}$Zr & 40 & 210 & 46.4 & $1.00\cdot 10^{12}$ &0.19 \\
 8 & $^{320}$Zr & 40 & 280 & 44.4 & $1.47\cdot 10^{12}$ &0.23 \\
 7 & $^{500}$Zr & 40 & 460 & 42.2 & $2.66\cdot 10^{12}$ &0.31 \\
 6 & $^{950}$Sn & 50 & 900 & 39.3 & $6.24\cdot 10^{12}$ &0.43 \\
 5 & $^{1100}$Sn & 50 & 1050 & 35.7 & $9.65\cdot 10^{12}$ &0.51 \\
 4 & $^{1350}$Sn & 50 & 1300 & 33.0 & $1.49\cdot 10^{13}$ &0.60\\
 3 & $^{1800}$Sn & 50 & 1750 & 27.6 & $3.41\cdot 10^{13}$ &0.80 \\
 2 & $^{1500}$Zr & 40 & 1460 & 19.6 & $7.94\cdot 10^{13}$ &1.08 \\
 1 & $^{982}$Ge & 32 & 950 & 14.4 & $1.32\cdot 10^{14}$ &1.33 \\
\hline
\end{tabular}
\caption{
The WS cells representing different density regions of the inner crust. 
The particle numbers Z,N, the WS-cell radii
$R_{WS}$ and the baryonic density $\rho_b$ have been taken from previous calculations \cite{Negele_1973}. 
$k_{F,n}$ is the Fermi momentum corresponding to the density of the outer neutron gas, as
computed in this work.}
\label{tabWS}
\end{center}
\end{table}

The Skyrme functional SLy4 has been used throughout this work, 
except for Figs.~\ref{density_instabilities_1500Zr_compare} and \ref{pairing_gaps_compare},
where a comparison with the functionals SkM* \cite{Chabanat_1997,Chabanat_1998} and MHF \cite{Hebeler_2009} is shown.

We consider three different two-body pairing interactions: (i) a density-dependent 
contact interaction; (ii) the finite-range Gogny D1 interaction; (iii) low-momentum
realistic interactions ($\vlowk$).

We restrict the Gogny D1 and the $\vlowk$ pairing interactions 
to the $\partialwave{1}{S}{0}$ partial wave.
Here we use the standard notation $^{2S+1}l_{J_{\rm rel}}$, with the Cooper-pair total spin $S$,
the relative orbital angular momentum $l$ and the relative total angular momentum
${\bm J}_{\rm rel} = {\bm l} + {\bm S}$.
Higher partial waves can also contribute to the superfluidity in finite nuclei, with P waves
giving a $\approx 15\%$ quenching of the S-wave pairing gaps \cite{Baroni_2010}.
This contribution could be even smaller in the inner crust of neutron stars, 
where the states close to the Fermi surface are in the
continuum and the center of mass of the Cooper pairs plays a less important role.

We now give a detailed description of the pairing interactions that we used.
\begin{itemize}
\item[(i)]
The two-body contact force DDDI 
between particles at positions $\mathbf{r_1}$ and $\mathbf{r_2}$ reads \cite{Bertsch_1991}
\begin{eqnarray}\label{pairing_int_contact}
\qquad \quad v(\mathbf{r}_{1},\mathbf{r}_{2})=V_{0}\left[ 1- \eta \left( \frac{\rho_b\left(
\frac{\mathbf{r}_{1}+\mathbf{r}_{2}}{2}\right)}{\rho_0}\right)^{\alpha}\right]
\delta(\mathbf{r}_{1}-\mathbf{r}_{2}), & & \nonumber \\
& &
\end{eqnarray} 
with $V_{0}=-430.0$ MeV fm$^3$, $\eta=0.7$, $\alpha=0.45$, $\rho_0=0.16 \fmicube$. 
We use a cutoff of $60\mev$ on the quasiparticle energy.
According to the literature \cite{Sandulescu_2004}, this parametrization is such that 
it approximately reproduces the Gogny D1 pairing gaps in
HFB calculations in INM.\\
\item[(ii)]
A separable pairing interaction 
that reproduces the $\partialwave{1}{S}{0}$ Gogny D1 pairing gap at the Fermi surface in INM
(see Refs.~\cite{Tian_2009a,Tian_2009b} for a detailed description)
\begin{eqnarray}\label{pairing_int_gogny}
  \qquad\quad
  \lefteqn{
           v(\mathbf{r}_1 ,\mathbf{r}_2 ,\mathbf{r}_1 ',\mathbf{r}_2 ') = 
          } \nonumber \\
   & & \qquad \gamma P(r)P(r')\delta (\mathbf{R} - \mathbf{R}')
       \frac{1}{2}(1-P^\sigma). \nonumber \\
& &
\end{eqnarray}
The operator $\frac{1}{2}(1-P^\sigma)$ restricts the interaction to total spin $S=0$.
$\mathbf{R}=(\mathbf{r_1}+\mathbf{r_2})/2$ is the center of mass of the two
interacting particles and $\mathbf{r}=\mathbf{r_1}-\mathbf{r_2}$ is their mutual distance.
Strength and form factor are $\gamma=-738\mev\fmicube$ and 
$P(r)= 1/(4\pi a^2 )^{3/2} \exp ( - r^2 /(4a^2 ))$, where $a=0.636$.
\item[(iii)]
A rank-3 separable interaction \cite{Duguet_2004, Lesisnki_2008, Lesinski_2009} of the form 
\begin{eqnarray}\label{pairing_int_vlowk}
  \qquad \quad
  v(\mathbf{r}_1 ,\mathbf{r}_2 ,\mathbf{r}_1 ',\mathbf{r}_2 ') = \sum\limits_{\beta  = 1}^3 {\lambda _\beta  }
  G_\beta(r)G_\beta(r')\delta (\mathbf{R} - \mathbf{R}') & & \nonumber \\
  & &
\end{eqnarray}
is used in this work to reproduce to high precision 
the $\partialwave{1}{S}{0}$ matrix elements of the low-momentum nucleon-nucleon interactions $\vlowk$
obtained from the Argonne potential AV18 . 
The latter has been RG evolved to a low-momentum cutoff $\Lambda$ using a smooth regulator $n_{exp}=6$.
The results shown in this paper are obtained with a separable force corresponding
to a $\Lambda=2.5\fmi$ low-momentum interaction. 
Our results are cutoff independent to a good approximation,
with the neutron pairing gaps changing of at most $30\kev$ and $100\kev$ for INM and 
for the WS cells respectively, when the cutoff $\Lambda$ ranges between $1.8\fmi$ and $4.0\fmi$.
The $G_\beta(r)$ form factors are a product of a Gaussian and a Hermite polynomial.
\end{itemize}

Both neutrons and protons are found to be superfluid in the WS cells of Table~(\ref{tabWS}),
with the proton gaps comparable to the neutron ones. 
However, in this work we discuss only neutron superfluidity.
We dropped the Coulomb term in the proton-proton pairing channel.
In a few cells (i.e., $\elemA{1350}{Sn}$, $\elemA{1800}{Sn}$ and $\elemA{1500}{Zr}$ ), 
we checked that the inclusion of the Coulomb term in the Gogny D1 pairing interaction leads to a quenching
of the proton pairing gaps between $20\%$ and $30\%$. 
This is in agreement with recent studies on finite nuclei \cite{Nakada_2011}. 
The neutron properties are not sensitively affected and 
discussion and conclusions are not changed by the inclusion of the Coulomb term,
as neutron LCS gaps are affected at the level of $1 \kev$. 

\subsection{Infinite neutron matter}

To study the effect of the proton clusters,
the superfluid properties of the inner crust are compared with those of the INM.
For a given neutron density $\rho_n$ in INM, the HFB gap and number equations
have to be solved simultaneously
\begin{equation}\label{gapeqINM}
\Delta_n(k)=-\frac{1}{2}\int \frac{d^3k'}{(2\pi)^3}v(k-k')\frac{\Delta_n(k')}{E_n(k')}
\end{equation}
\begin{equation}\label{numbeqINM}
\rho_n=\frac{1}{2\pi^2}\int dk \; k^2
                \left[
                 1-\frac{\varepsilon_n(k)-\mu_n}{E_n(k)}
                \right] .
\end{equation}
$\mu_n$ is the neutron chemical potential and $E_n(k)=\sqrt{(\varepsilon_n(k)-\mu_n)^2+\Delta_n(k)^2}$ 
is the quasiparticle energy, while the single-particle energy $\varepsilon_n(k)$ is given by
the sum of the kinetic energy and the Hartree-Fock potential $\bar{U}^n_{HF}$
\begin{equation}\label{spINM}
  \varepsilon_n(k)=\frac{\hbar^2k^2}{2m_n^*} + \bar{U}^n_{HF}(k) .
\end{equation}
We use the Skyrme neutron effective mass $m_n^*$.
The number equation (cf. Eq.~(\ref{numbeqINM})) provides the relation between the density 
and the chemical potential $\mu_n$. In the limit of weak coupling, where $\Delta_n << \varepsilon_{F,n}$,
the chemical potential can be approximated by the Fermi energy $\varepsilon_{F,n}=\frac{\hbar^2 k_{F,n}^2}{2m_n^*}$,
with $k_{F,n}=(3\pi^2\rho_n)^{1/3}$. This approximation somewhat holds already at
$k_{F,n} \approx 0.2 \fmi$, and we are left with solving only the gap equation (cf. Eq.~(\ref{gapeqINM})). 
We check the validity of the above approximation a posteriori, by comparing
the solution of Eq.~(\ref{gapeqINM}) with that obtained in
spherical-box calculations of homogeneous neutron matter, 
where both gap and number equations are solved simultaneously.
The good agreement between the two methods (see Sect.~\ref{Sect:results_pairing_gaps})
supports our results.
Since the neutron HF potential $\bar{U}^n_{HF}$ is constant and the single-particle energies are taken from
the Fermi level, we also adopt the approximation 
$\varepsilon_n(k)=\frac{\hbar^2k^2}{2m_n^*}$.

\section{Results}\label{Sect:results}

\begin{figure*}
\begin{center}
\includegraphics[clip=,width=0.75\textwidth,angle=0]{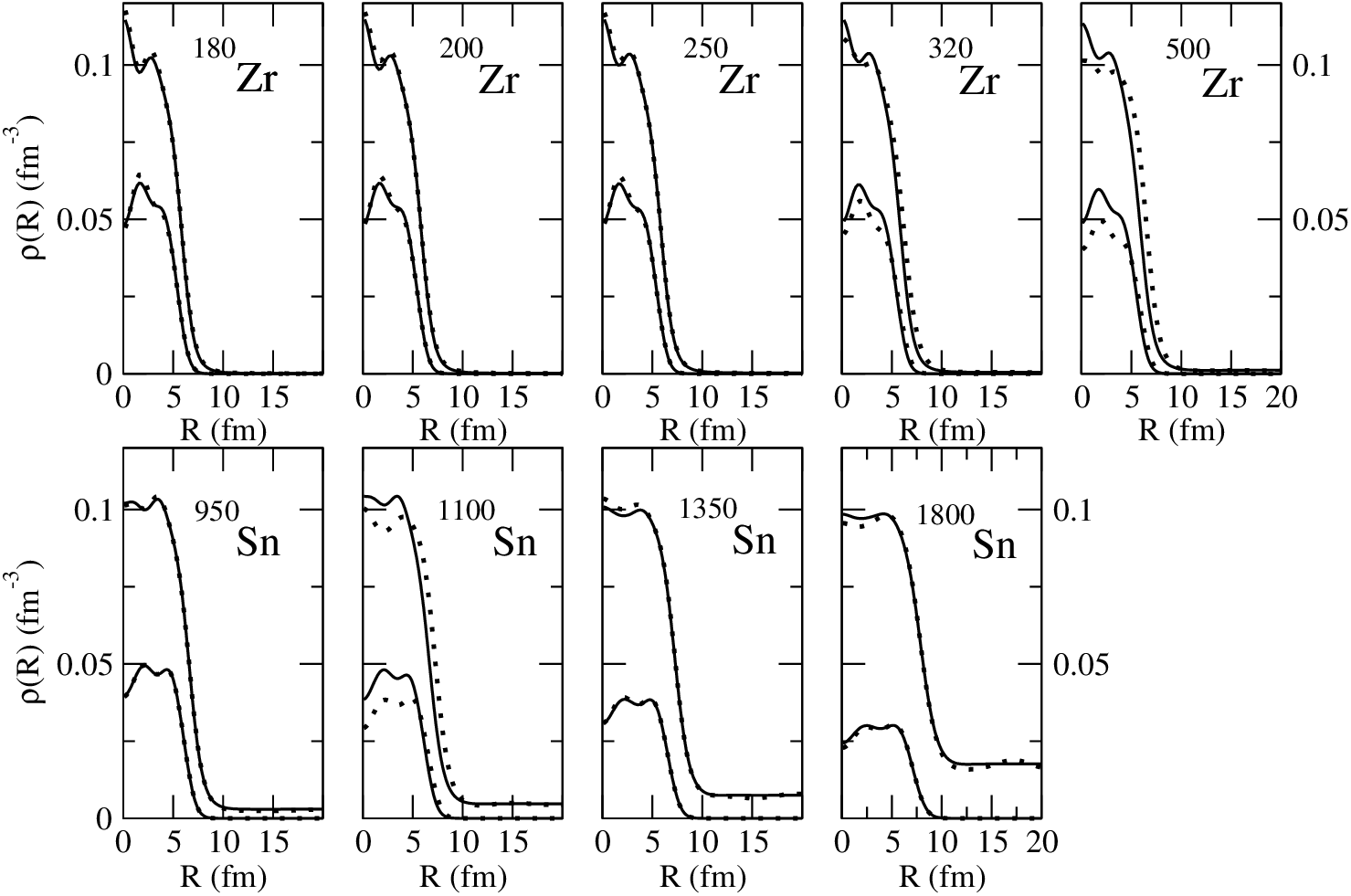}
\end{center}
\caption{HF (dotted lines) and HFB (solid lines) 
neutron and proton densities for the inner-crust regions 11 through 3 (see Table~(\ref{tabWS})).
The HFB results have been obtained using the $\vlowk$ pairing interaction.
The neutron (proton) density corresponds to the upper (lower) curves of each panel.
} 
\label{densities_1}
\end{figure*}

\begin{figure}
\begin{center}
\includegraphics[clip=,width=0.40\textwidth,angle=0]{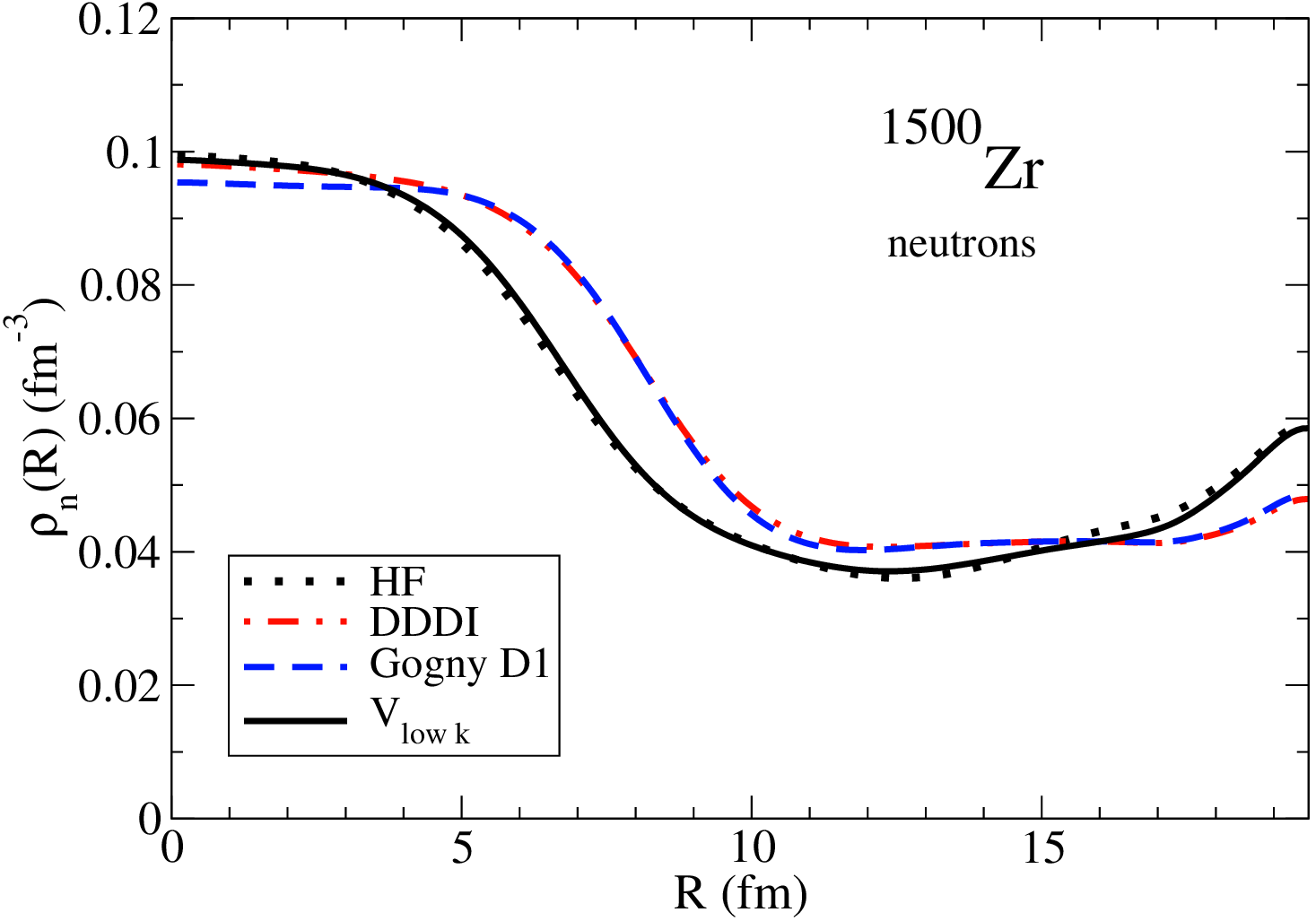}
\includegraphics[clip=,width=0.40\textwidth,angle=0]{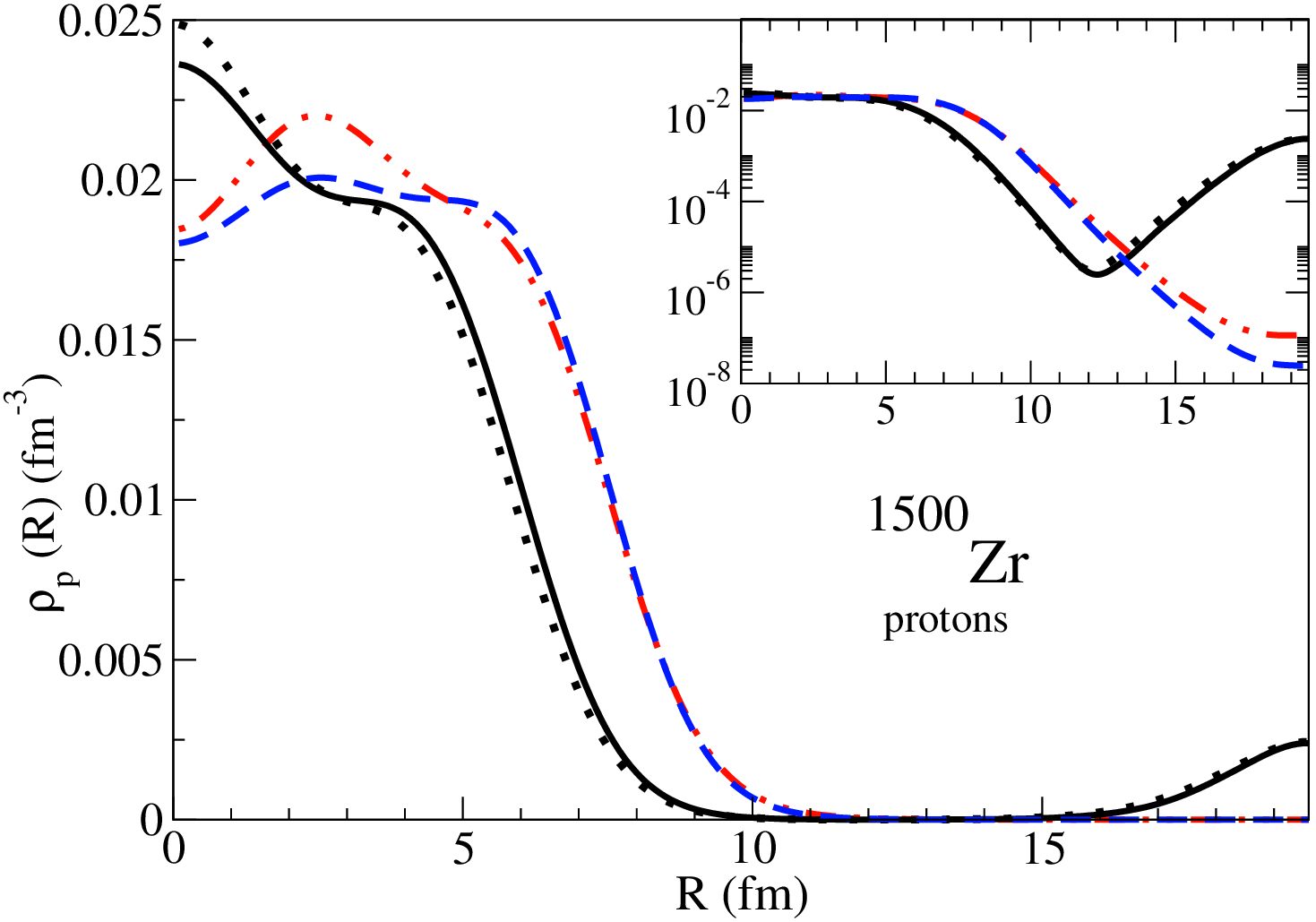}
\end{center}
\caption{(Color online) Neutron (top panel) and proton (bottom panel) densities 
for the $\elemA{1500}{Zr}$ WS cell.
The inset shows the proton densities in a semilogarithmic scale.} 
\label{density_instabilities_1500Zr}
\end{figure}

\begin{figure}
\begin{center}
\includegraphics[clip=,width=0.40\textwidth,angle=0]{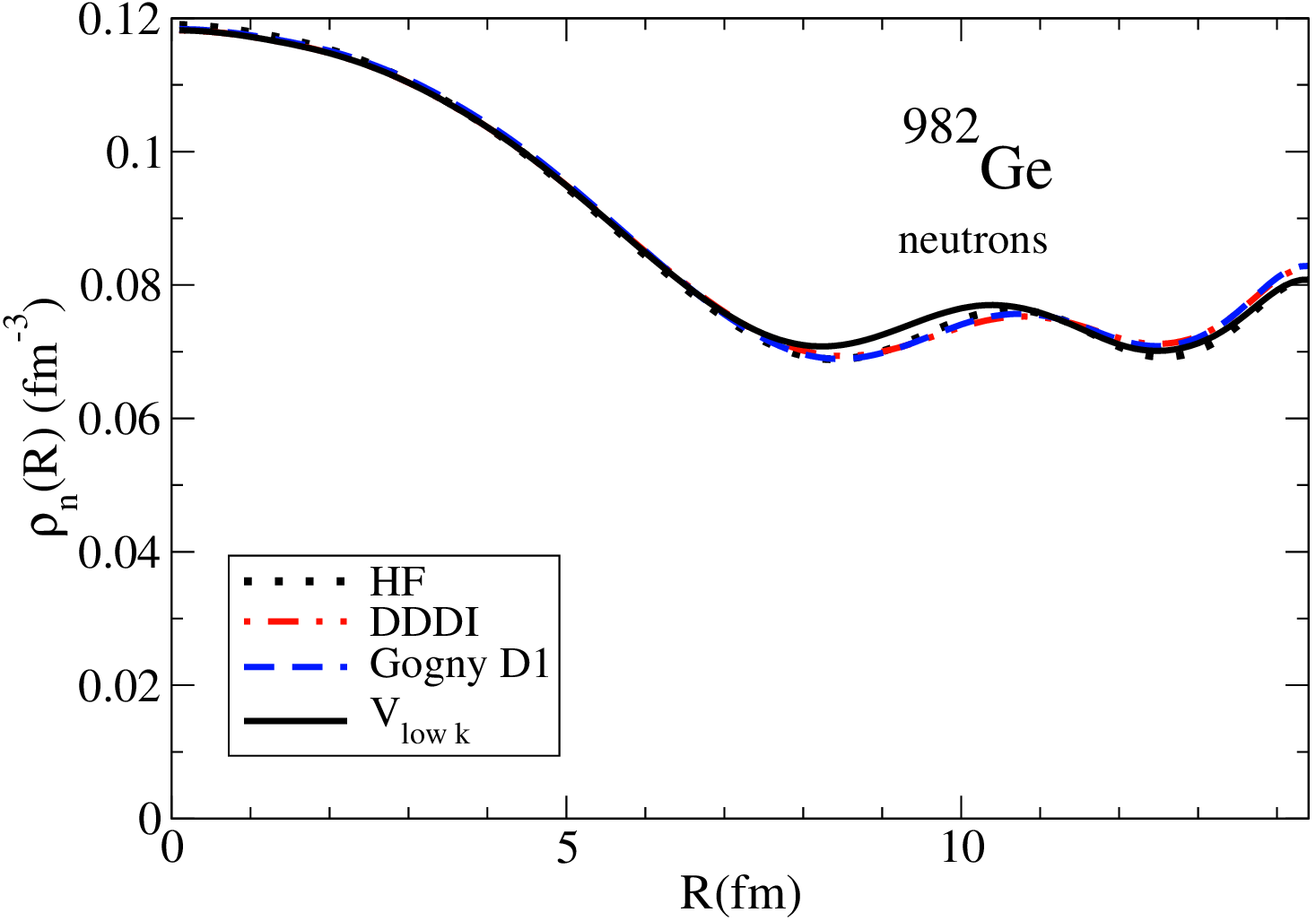}
\includegraphics[clip=,width=0.40\textwidth,angle=0]{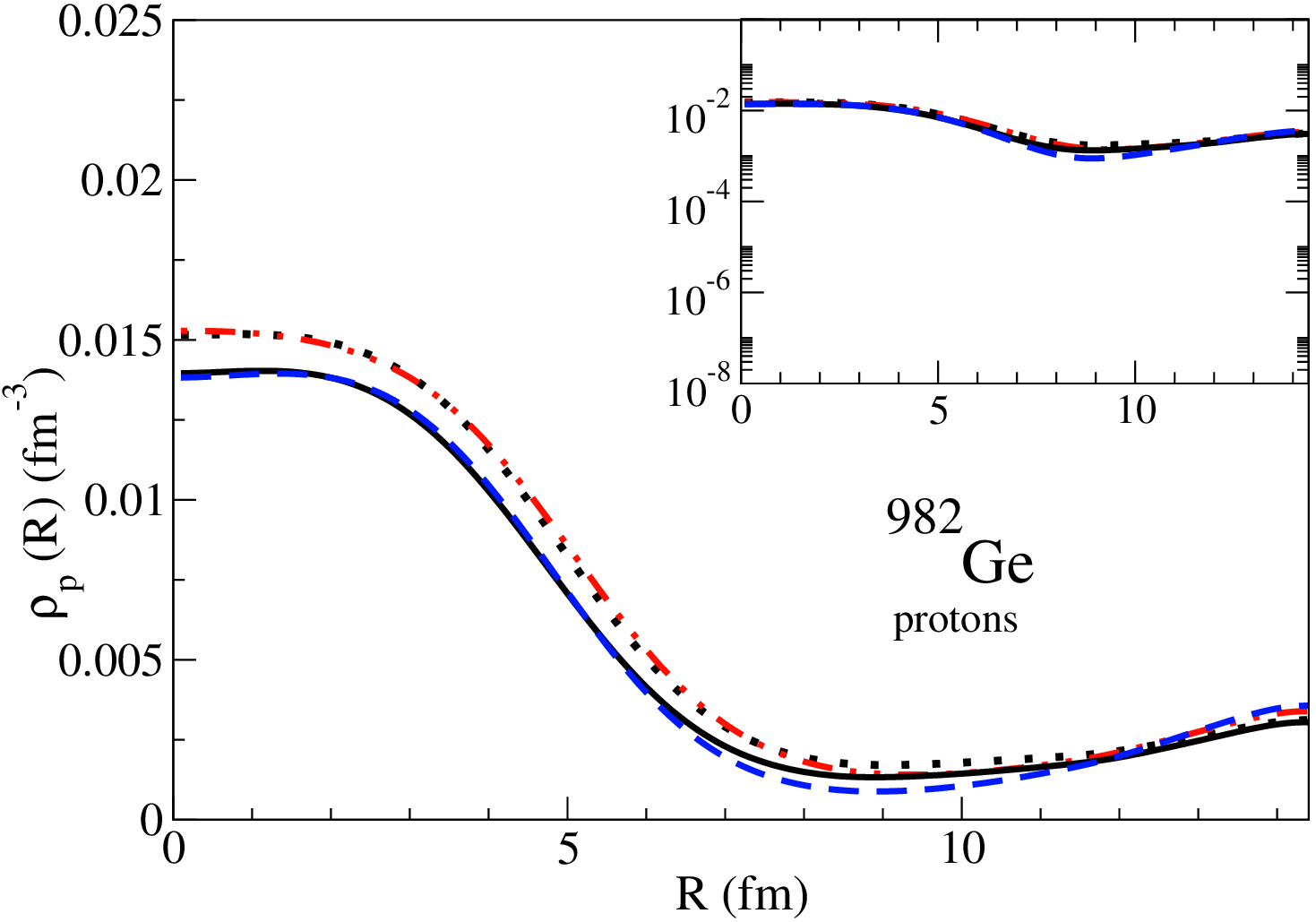}
\end{center}
\caption{(Color online) Same as Fig.~\ref{density_instabilities_1500Zr}, for 
the $\elemA{982}{Ge}$ WS cell.} 
\label{density_instabilities_982Ge}
\end{figure}

\begin{figure}
\begin{center}
\includegraphics[clip=,width=0.40\textwidth,angle=0]{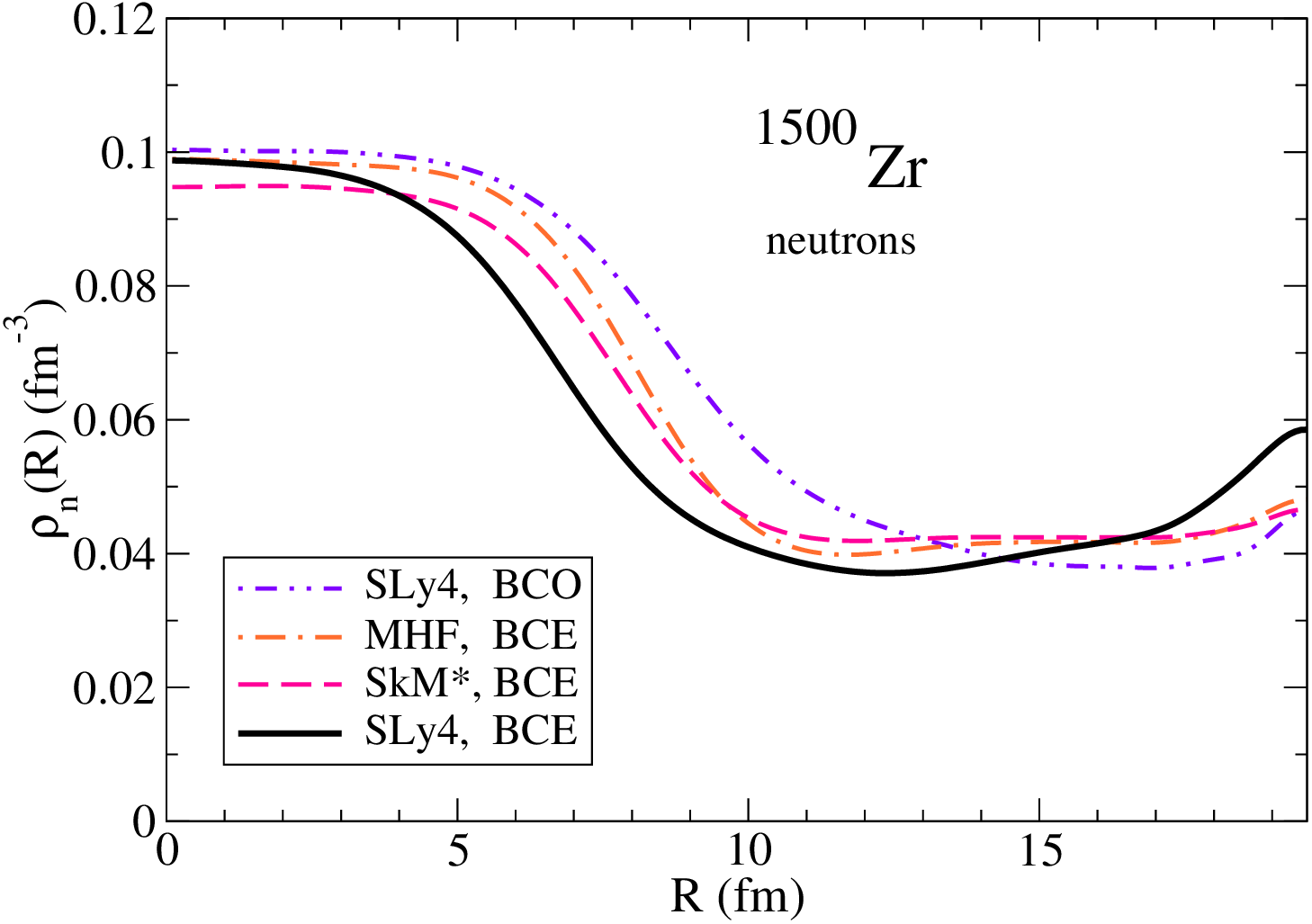}
\includegraphics[clip=,width=0.40\textwidth,angle=0]{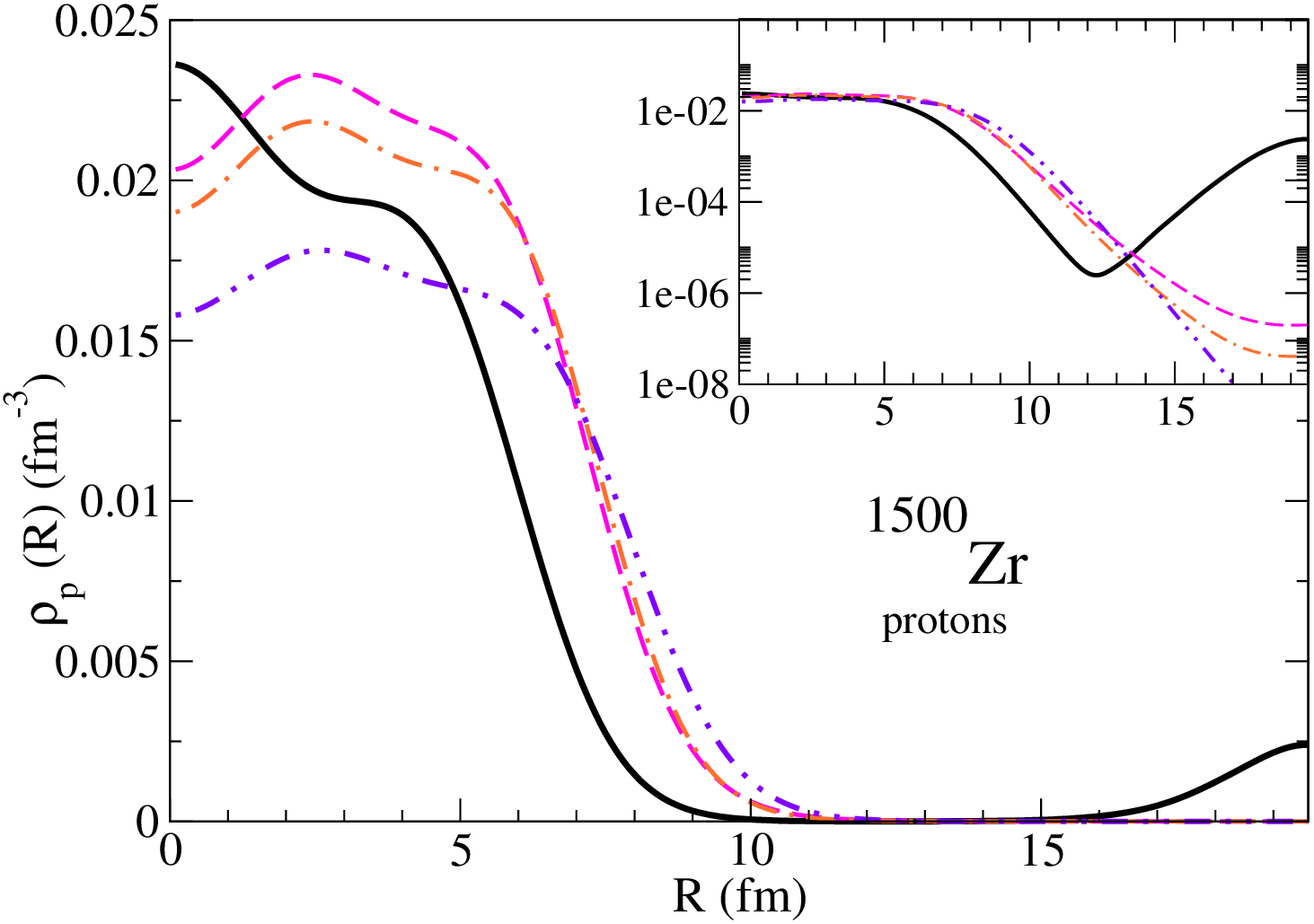}
\end{center}
\caption{(Color online) Neutron (top panel) and proton (bottom panel) densities 
for the $\elemA{1500}{Zr}$ WS cell, using 
the $\vlowk$ pairing interaction on top of different Skyrme functionals and 
for different boundary conditions.
The inset shows the proton densities in a semilogarithmic scale.} 
\label{density_instabilities_1500Zr_compare}
\end{figure}

\subsection{Density profiles and limits of the WS approximation}\label{Sect:results_density}

HF and HFB neutron and proton densities for the inner-crust regions 11 through 3 
(cf. Table~(\ref{tabWS})) 
are shown in Fig.~\ref{densities_1}.
In these regions, the HFB density profiles obtained with the three pairing interactions are almost 
on top of each other.
The density of the outer neutrons gradually raises as one goes deeper and deeper into the inner crust, 
with the innermost WS cell $\elemA{1800}{Sn}$ having an outer neutron 
density $\rho_n\approx 0.02\fmicube$.
The proton clusters have a radial extension that ranges from $5\fm$ for the outermost cell 
(i.e.,$\elemA{180}{Zr}$) to about $7.5\fm$ for $\elemA{1800}{Sn}$. 
The proton density above $10\fm$ is negligible for all regions 11 through 3.

At higher baryonic densities ($\rho_b \ge 0.25 \rho_0$), 
the mean-field proton spatial distribution becomes
unstable, as shown in Figs.~\ref{density_instabilities_1500Zr} and \ref{density_instabilities_982Ge}
for the two high-density cells $\elemA{1500}{Zr}$ and $\elemA{982}{Ge}$.
The proton density does not correspond to that of a proton cluster, as a
non-negligible number of protons are sitting at the edge of the $\elemA{1500}{Zr}$ cell, 
while the protons are spread out over the whole $\elemA{982}{Ge}$ cell.

For the $\elemA{1500}{Zr}$ cell, the proton density also depends on the strength of the pairing 
interaction (see bottom panel of Fig.~\ref{density_instabilities_1500Zr}). For HF calculations 
(i.e., zero pairing strength) 
and for HFB calculations with $\vlowk$ in the pairing channel, the box effect is much stronger than 
for calculations with the Gogny D1 and the DDDI pairing forces.

These instabilities are sensitive to the Skyrme functional used and to the 
boundary conditions as well. Neutron and proton HFB densities for the $\elemA{1500}{Zr}$ cell
are shown in Fig.~\ref{density_instabilities_1500Zr_compare} for calculations using the $\vlowk$
pairing interaction on top of a mean field built with SLy4, SkM* and MHF Skyrme functionals.
The results obtained with the BCO boundary conditions are also shown for the SLy4 functional
in the same figure.
The proton instability in $\elemA{1500}{Zr}$ is present only for SLy4 with BCE boundary conditions.
All calculations for the higher-density cell $\elemA{982}{Ge}$ 
give results very similar to each other, with the protons spread out over the 
entire volume of the WS cell.

We conclude that solving the HFB equations in a WS cell of densities $\rho_b \ge 0.25 \rho_0$
not always leads to a reliable solution. The method starts to be unstable at baryonic densities 
$\rho_b \approx 8 \cdot 10^{13}$ g $\cdot$ cm$^{-3} \approx 0.25 \rho_0$ and it definitely breaks down at 
$\rho_b \approx 10^{14}$ g $\cdot$ cm$^{-3} \approx 0.35 \rho_0$.

We believe that the WS approximation should not be used whereas these instabilities occur.
Hence, in the following we show results only for the calculations where the HFB solution is stable.

\subsection{Pairing gaps}\label{Sect:results_pairing_gaps}

\begin{figure*}
\begin{center}
\includegraphics[clip=,width=0.45\textwidth]{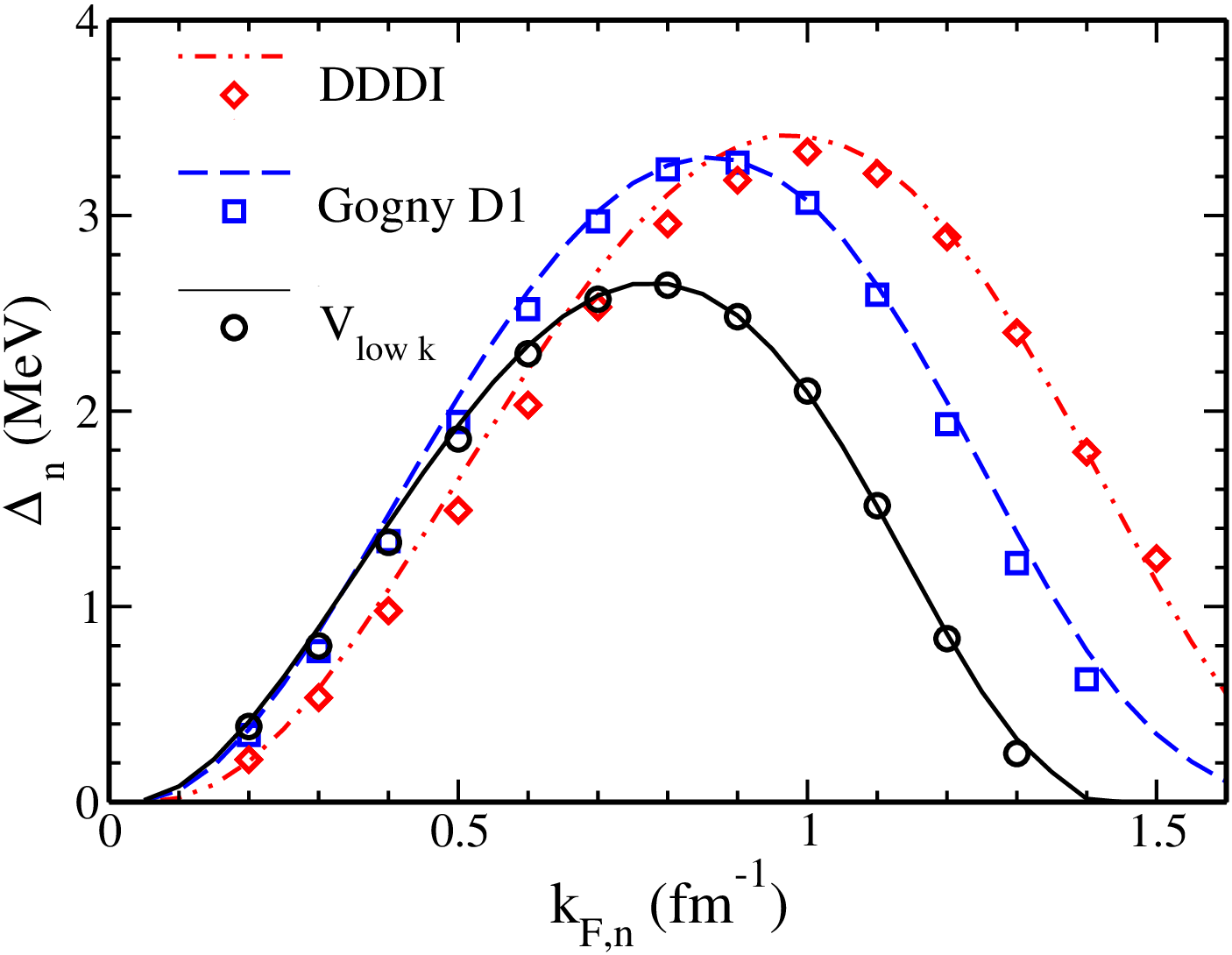}
\includegraphics[clip=,width=0.45\textwidth,angle=0]{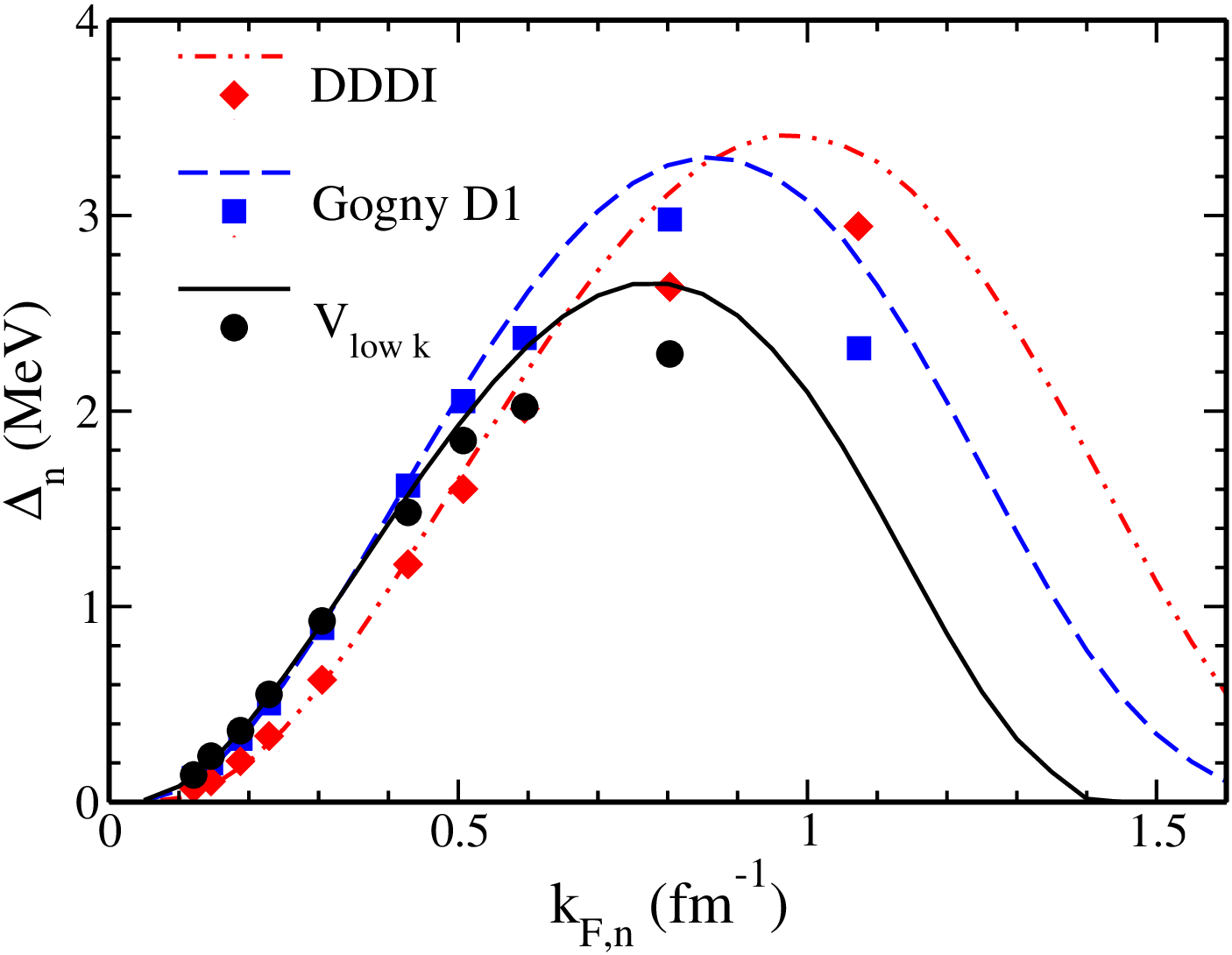}
\end{center}
\caption{
(Color online) Pairing gaps obtained with the SLy4 Skyrme functional and the three pairing interactions in the legend. 
The lines represent the pairing gaps in infinite neutron matter and correspond to the solution $\Delta_n(k_{F,n})$ 
of Eq.~(\ref{gapeqINM}). 
Diamonds, squares and circles correspond to box calculations with DDDI, Gogny D1 and $\vlowk$ pairing interactions, 
respectively.
The open points in the left panel are the LCS pairing gaps in infinite neutron matter, 
obtained from spherical-box HFB calculations with no protons (see text).
The solid dots in the right panel are the LCS neutron pairing gaps in the inner crust, for the WS cells 
11 through 2 (see Table~(\ref{tabWS})).
}
\label{pairing_gaps}
\end{figure*}

It is useful to compare the superfluid properties of the inner crust to 
those of the INM.
The pairing gaps at the Fermi surface are shown in Fig.~\ref{pairing_gaps} for INM and the inner crust.

\begin{figure*}
\includegraphics[clip=,width=0.45\textwidth,angle=0]{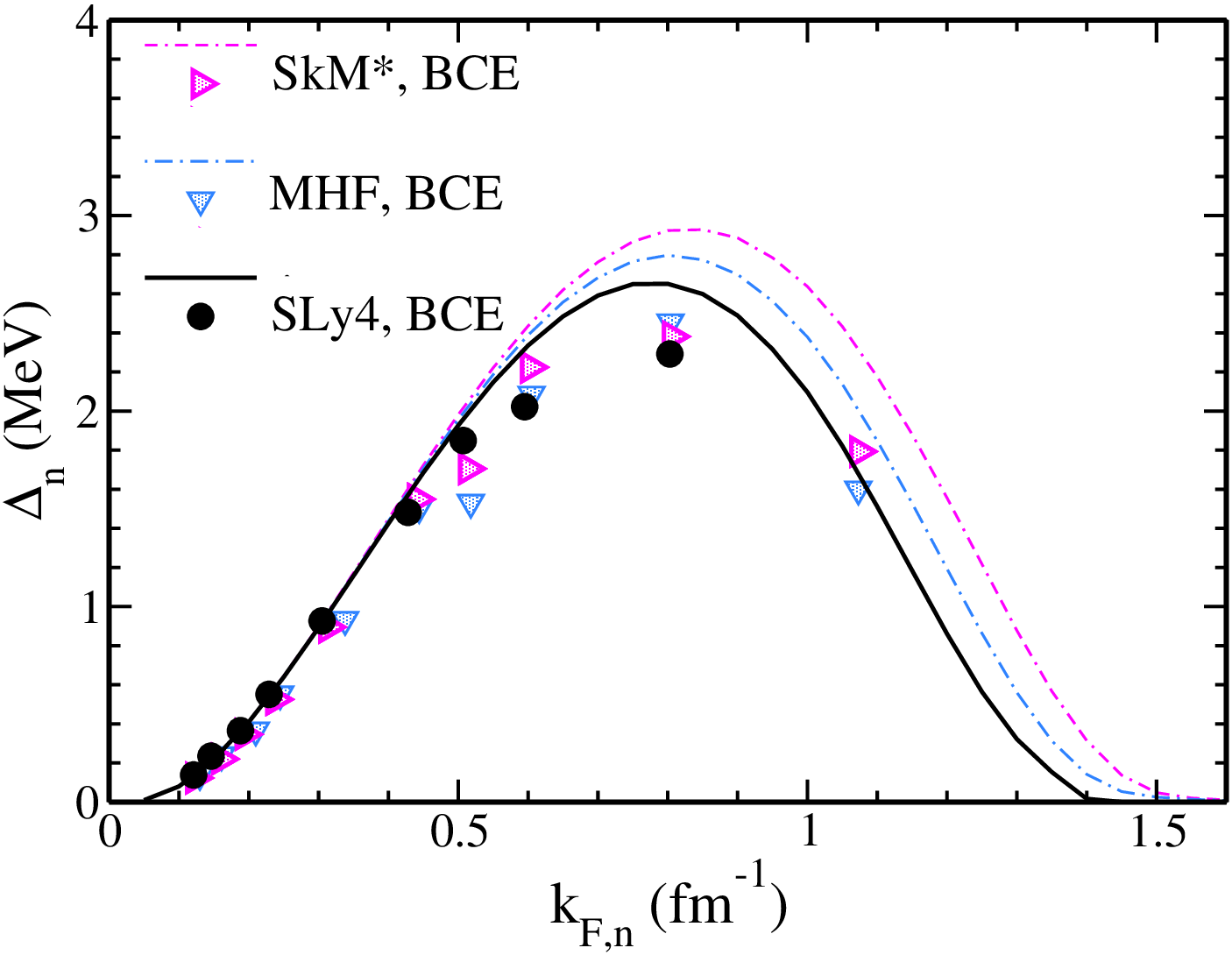}
\includegraphics[clip=,width=0.45\textwidth,angle=0]{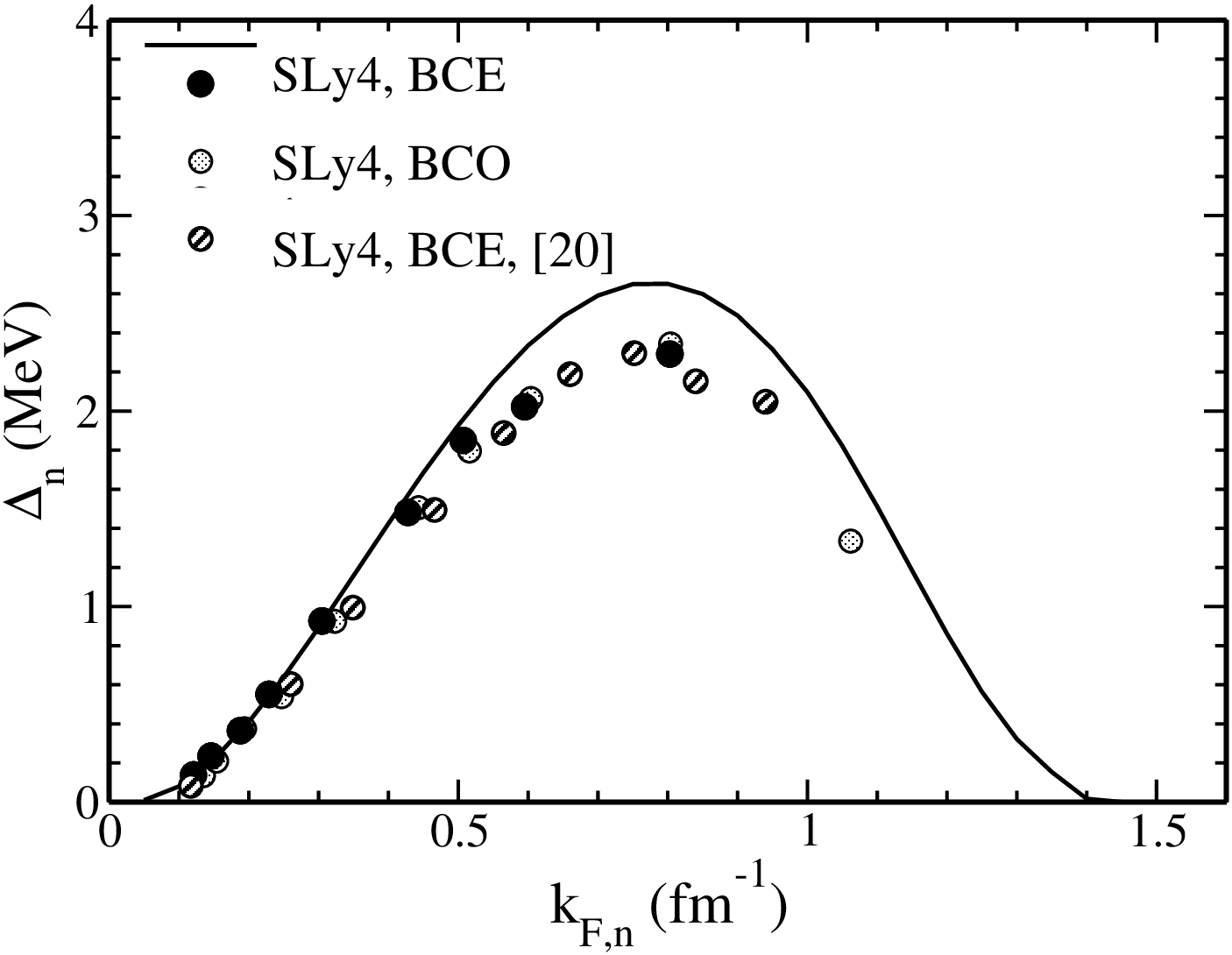}
\caption{(Color online) 
Dependence of the pairing gap on the Skyrme functional, on the boundary conditions and on the 
WS-cell ($R_{WS}, Z$) parameters. 
All results are obtained with the $\vlowk$ pairing interaction on top of different Skyrme functionals.
The lines represent the pairing gaps in infinite neutron matter and correspond to the solution 
$\Delta_n(k_{F,n})$ of Eq.~(\ref{gapeqINM}).
The points represent the pairing gaps for the inner crust, obtained in box calculations.
The solid line and the solid dots are the same as in Fig.~\ref{pairing_gaps} and are obtained
with the SLy4 Skyrme functional and BCE boundary conditions.
Left panel: HFB solutions obtained with SkM*, MHF and SLy4 Skyrme functionals and with BCE
boundary conditions.
Right panel: results are obtained with the SLy4 Skyrme functional. Solid and graded dots correspond to the two
different BCE and BCO boundary conditions, respectively. The striped dots are the HFB solutions
for the WS-cell ($R_{WS}, Z$) parameters of Ref.~\cite{Baldo_2005}.}
\label{pairing_gaps_compare}
\end{figure*}

The lines represent the solutions $\Delta_n(k_{F,n})$ of Eq.~(\ref{gapeqINM}) in infinite neutron matter. 
The open points are the LCS pairing gaps obtained from HFB calculations in spherical boxes of radius 
$R_{WS}=40\fm$, with no protons (Z=0) and with neutron number given by
$N=k_{F,n}^3/(3\pi^2)\cdot 4\pi R_{WS}^3 / 3$. The solid points are the 
LCS neutron pairing gaps in the inner crust regions 11 through 2 (see Table~(\ref{tabWS})).

As can be seen from the left panel of Fig.~\ref{pairing_gaps}, the agreement between
box and infinite-matter calculations is very good.
However, the INM results obtained with the three pairing interactions differ substantially 
from each other.
In particular, the gaps from the $\vlowk$ interactions are sensitively smaller than the others,
with a maximum of $2.6 \mev$ against $3.2 \mev$ for the effective pairing interactions.
At saturation, the Gogny pairing gap ($\approx 1.2\mev$) is much larger than the 
gap obtained with $\vlowk$ ($\approx 0.2\mev$). 
The results with the DDDI pairing force differ even more. 

On the one hand, because of its simplicity,
this type of contact pairing force is widely used in BCS and HFB
calculations in INM and in the inner crust. Its parameters are usually fitted to reproduce
given infinite-matter pairing gaps, and then used in inner-crust calculations.
Hence, many parameter sets and energy cutoffs have been used in the past. 
On the other hand, realistic pairing interactions are phase-shift-equivalent interactions,
they require no fitting procedures and allow us to connect the theory
to the basic nucleonic forces. The RG evolution to low-momentum and the separable
representation in the $\partialwave{1}{S}{0}$ channel make full HFB calculations
in the inner crust feasible and fast, even on a laptop computer.
We then advocate for adopting these realistic pairing interactions to get more reliable
results when applying microscopic theories to the inner crust of neutron stars. 
More than that, higher-order pairing correlations  
and other contributions to the pairing channel (i.e., three-body forces) 
represent necessary contributions \cite{Gori_2005, Gori_2004, Lesinski_2011} 
and the adoption of realistic nucleonic potentials allows one to include them consistently.

The effect of the proton clusters on the inner-crust pairing gaps is shown 
in the right panel of Fig.~\ref{pairing_gaps}.
The presence of the protons leads to a reduction of the pairing gap of at most
$10\%$ at the maximum of the curve. This effect is negligible below $k_{F,n}=0.5\fmi$.

The pairing gap at the Fermi surface depends on the Skyrme functional used,
which defines the level density.
Except for the highest-density regions close to the star core,
there is a very small dependence on the boundary conditions and on 
the particular WS-cell ($R_{WS}, Z$) parametrizations.
The pairing gap obtained with the $\vlowk$ pairing interaction on top of the Skyrme functionals
SkM*, MHF and SLy4 is shown in the left panel of Fig.~\ref{pairing_gaps_compare}.
The INM gap value for the three functionals reflects the different INM effective masses
$m_n^*/m_n$ at saturation density
(namely, $0.7531$, $0.8687$ and $0.997$ for SLy4, MHF and SkM$^*$ respectively).

The dependence of the pairing gap on the boundary conditions (namely, BCE and BCO) 
is shown in the right panel of 
Fig.~\ref{pairing_gaps_compare}. The gaps obtained using the $(R_{WS},Z)$ parameters from 
Ref.~\cite{Baldo_2005} are also shown in the same figure. All points lie in a narrow band.
One always has to keep in mind that either Refs.~\cite{Negele_1973, Baldo_2005} used
functionals and pairing interactions that differ from the Skyrme functionals
and from the pairing interactions used in this work.
One should use WS-cell $(R_{WS},Z)$ parameters obtained from an energy-minimization procedure
performed using the same functionals and pairing interactions later used to study the superfluid
properties of the system. While this goes beyond the goal of the present work,
we checked that even Ref.~\cite{Baldo_2005} predicts a $10 \%$ suppression of the pairing gap
at its maximum due to the presence of the protons.

\subsection{Spatial extension of the Cooper-pair}\label{Sect:results_cooperpair}

\begin{figure} 
\includegraphics[width=0.40\textwidth,clip=]{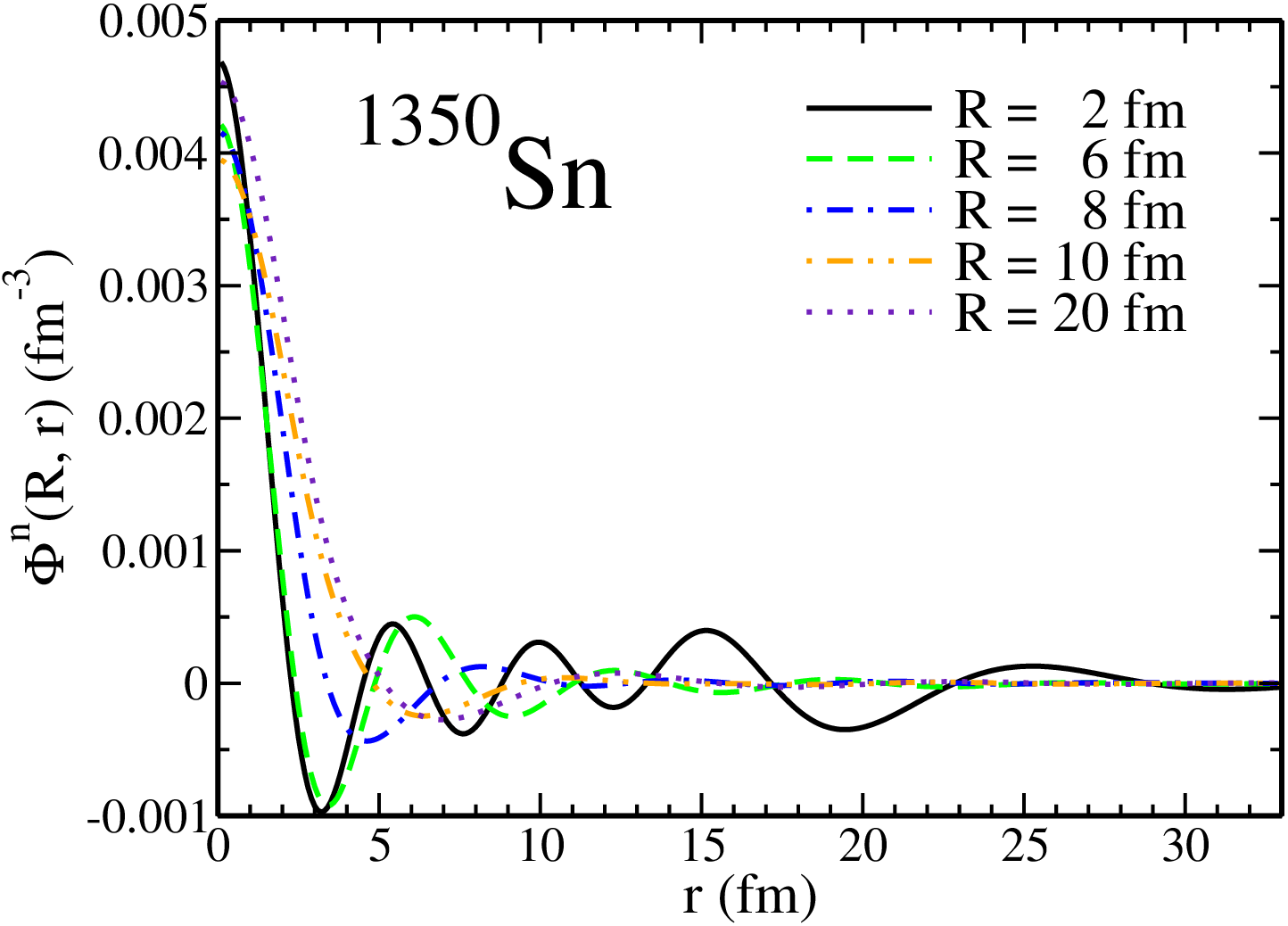}
\caption{(Color online) Cooper-pair wave function for $\elemA{1350}{Sn}$ as a function of the 
relative distance $r$ of the two nucleons of the pair.
Different curves correspond to different center-of-mass values R.}\label{cooper_Sn1350}
\end{figure}

\begin{figure}
\includegraphics[width=0.40\textwidth,clip=]{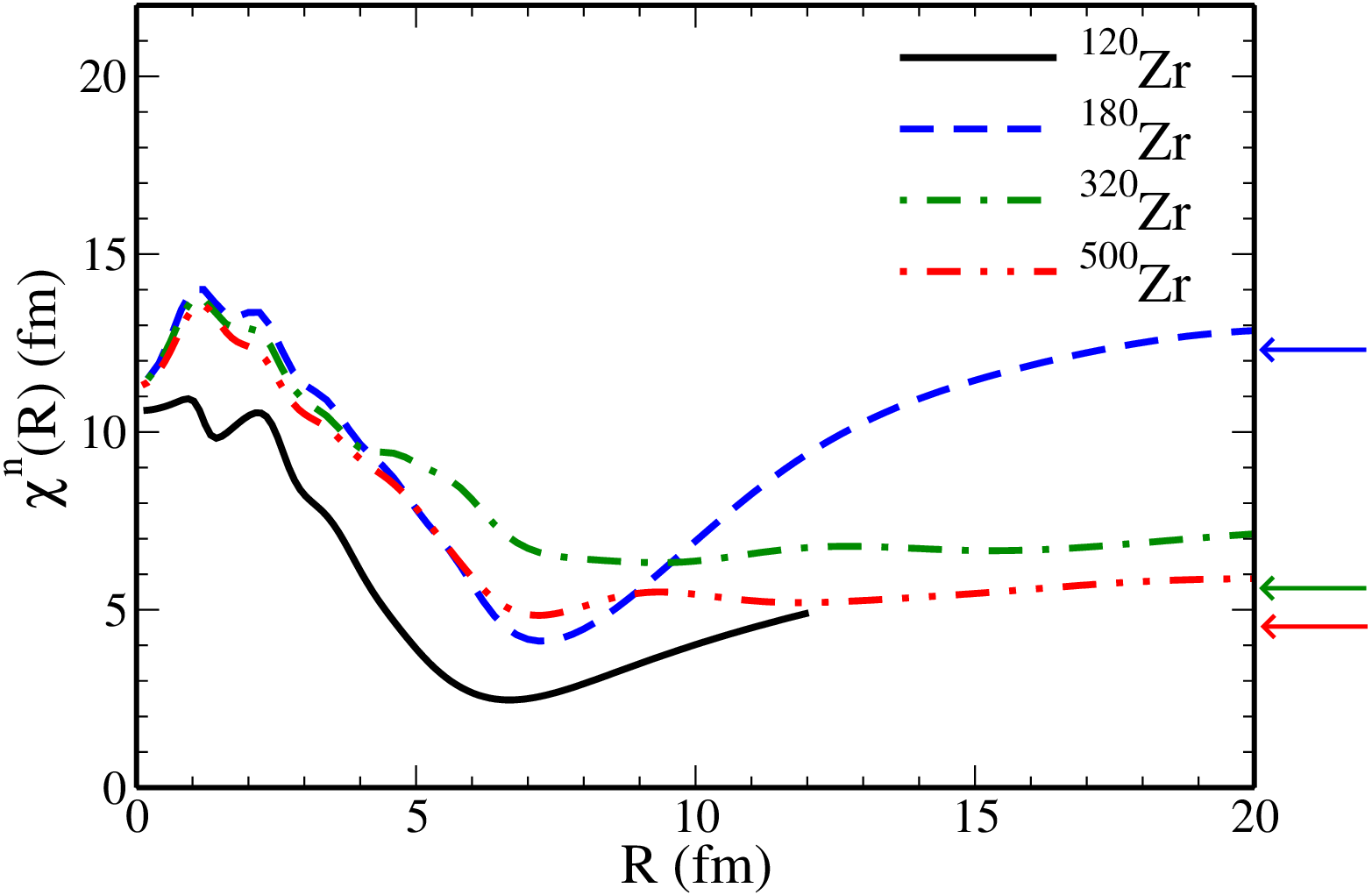}
\includegraphics[width=0.40\textwidth,clip=]{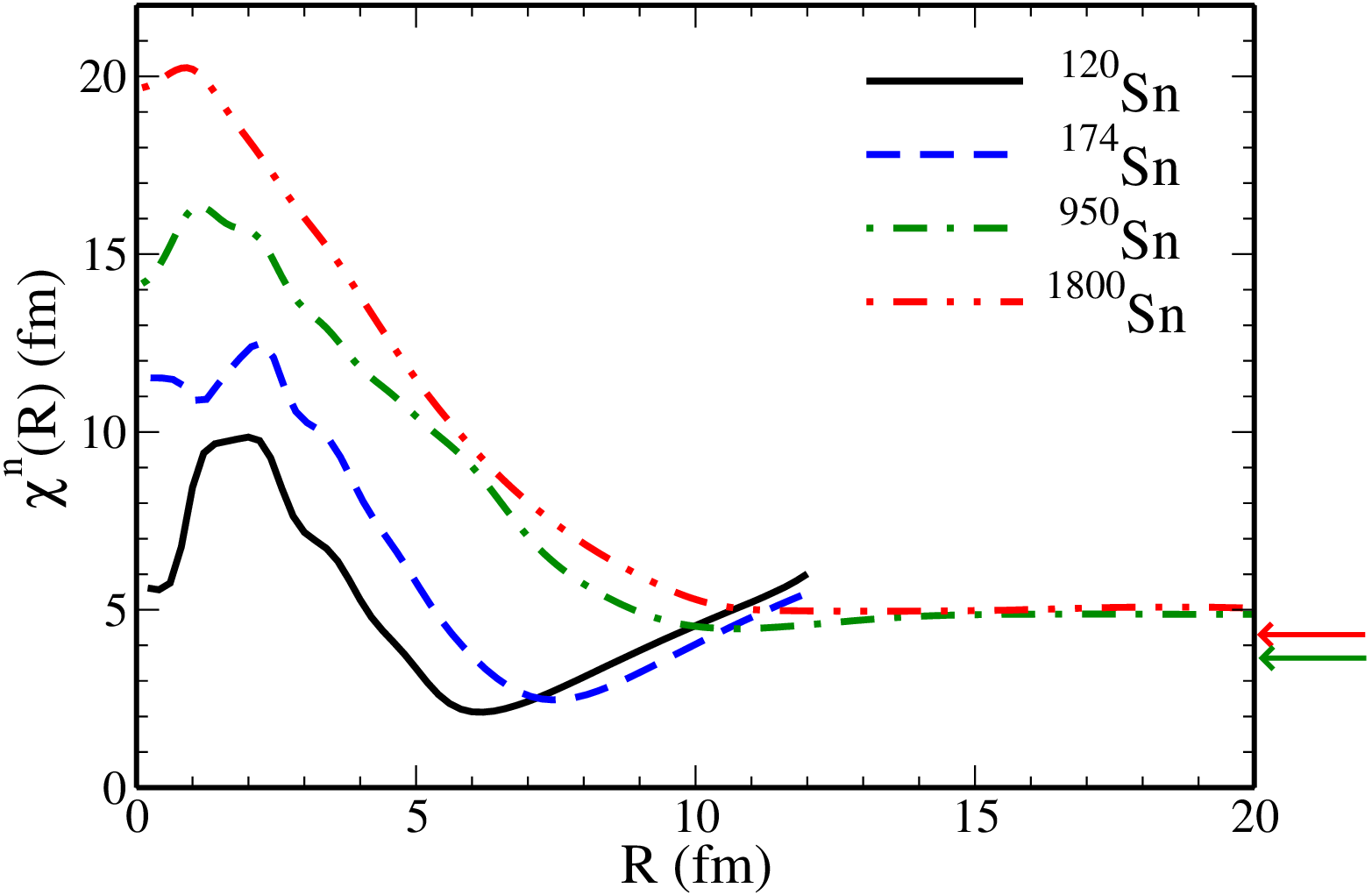}
\caption{(Color online) Neutron coherence length for different $Z=40$ (top panel) and $Z=50$ (bottom panel) 
systems, 
calculated using the SLy4 functional and the $\vlowk$ pairing interaction. 
The arrows on the right correspond to the INM coherence length taken at the Fermi momentum of
the outer neutron gas (see text for details).} 
\label{coherence_zrsn}
\end{figure}

\begin{figure*}
\includegraphics[width=0.40\textwidth,clip=]{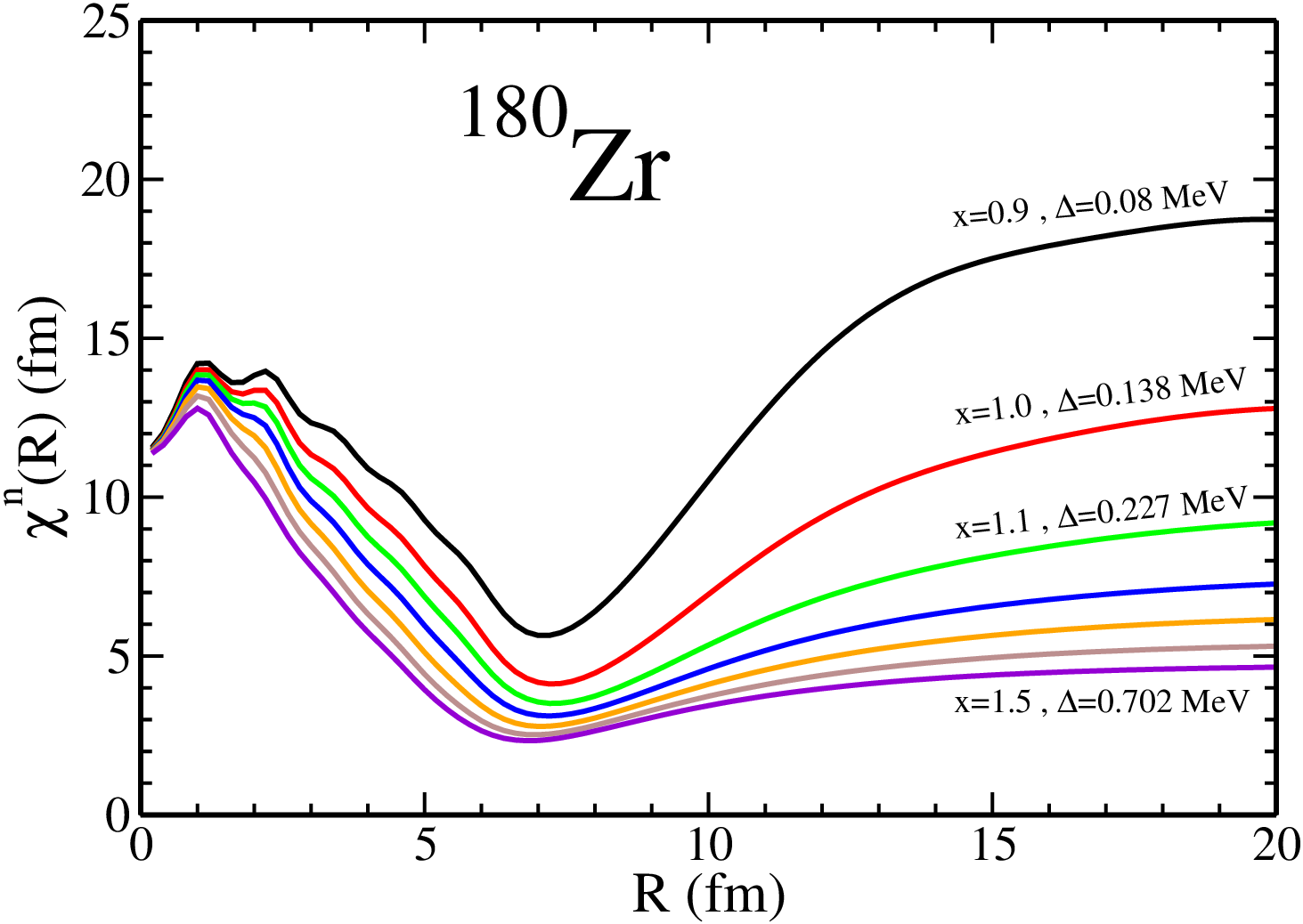}
\includegraphics[width=0.40\textwidth,clip=]{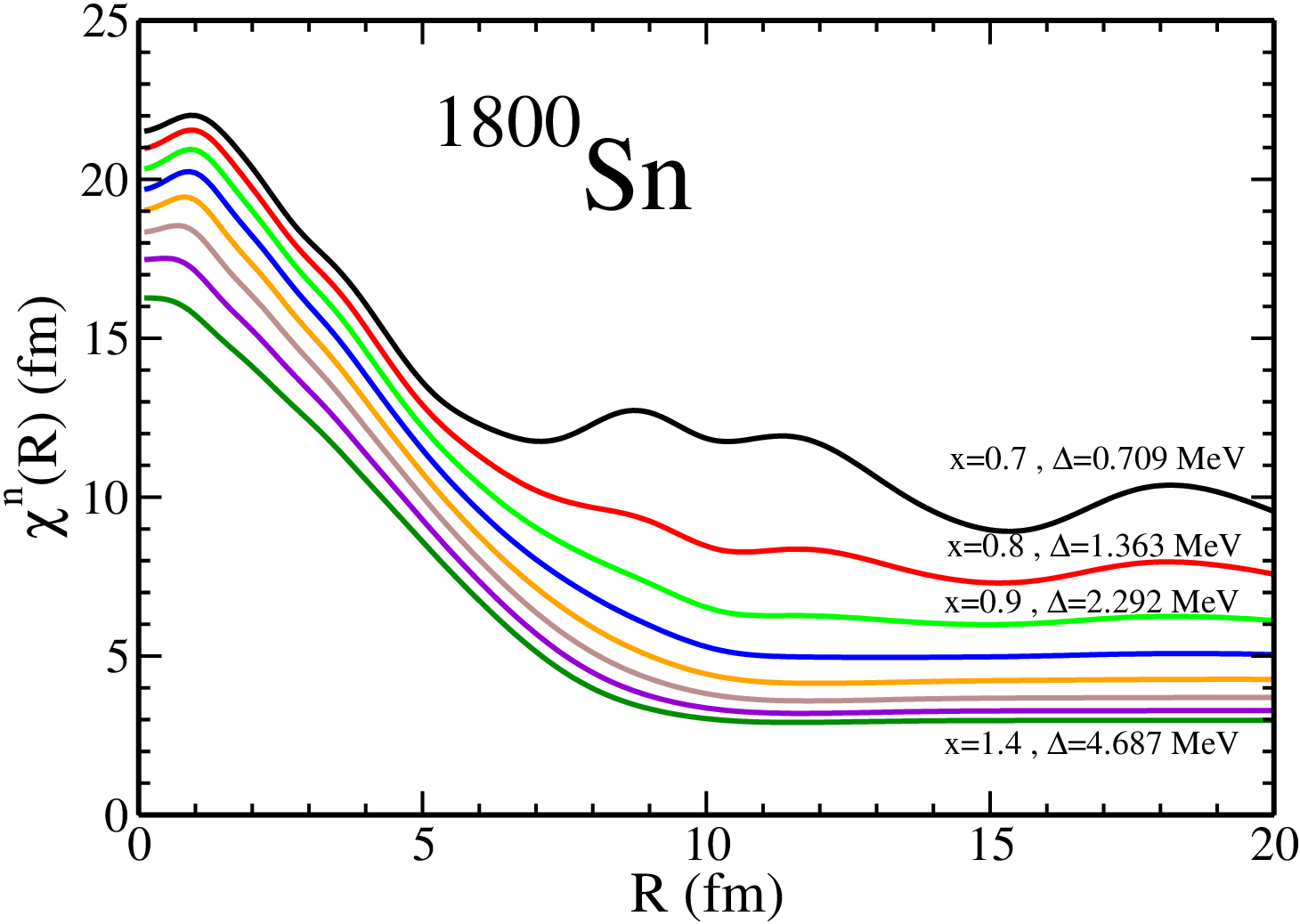}
\caption{(Color online) Neutron coherence length for $^{180}$Zr (left panel) and $^{1800}$Sn (right panel),
calculated using the $\vlowk$ pairing interaction rescaled by a factor $x$ (see text for details). 
From top to bottom, the curves correspond to an increasing strength of the pairing interaction.} 
\label{coherence_pairingstrength}
\end{figure*}

This section is dedicated to the study of the spatial properties of the Cooper pairs.
The Cooper-pair wave function is defined as an $UV$-weighted superposition of two-particle-state wave functions 
$[\phi(\mathbf{r_1})\phi(\mathbf{r_2})]_{00}$ coupled to total angular momentum $J=0$ and total spin $S=0$
\begin{equation}\label{a:density:r}
  \Phi^{q}(\mathbf{r}_1,\mathbf{r}_2) 
  =
  \sum_{inn'lj} \frac{2j+1}{2} {U^{i,q}_{nlj}V^{i,q}_{n'lj}} [\phi(\mathbf{r_1})\phi(\mathbf{r_2})]_{00}.
\end{equation}

In the following we will show the Cooper-pair wave function in the center-of-mass reference frame.
There is a weak dependence of $\Phi^q$ on the angle between  the Cooper-pair center of mass 
$\mathbf{R}$ and the relative position of the two nucleons
$\mathbf{r}$. Hence, an angular average can be performed
without loss of information. Results for the neutron wave function $\Phi^{n}(R,r)$
in the $\elemA{1350}{Sn}$ WS cell are shown in Fig.~\ref{cooper_Sn1350}. The dependence on the center of mass
vanishes outside the nucleus, for $R > 10 \fm$, recovering the asymptotic INM behaviour for
large $r$ values

\begin{equation}
\Phi^{n}_{INM} (r)\approx \frac{1}{r} K_{0}\left(r/\pi \chi^{n}_{P} \right) \sin (k_{F,n} r),
\end{equation}
where $K_{0}$ is the modified Bessel function and  $\chi^{n}_{P}$ is the 
neutron Pippard coherence length, defined as 

\begin{equation}\label{coher:pippard}
\chi_{P}^{n}=\frac{\hbar^{2} k_{F,n}}{m_n^{*}\pi \Delta^{n}_{F}}.
\end{equation}
The modified Bessel function, whose expression for large values of $r$ is
$K_{0}\left(r/\pi \chi^{n}_{P} \right)\approx (\chi^{n}_{P} /r)^{1/2}\exp[-(r/\pi\chi^{n}_{P} )]$, 
damps the oscillating behavior of the INM coherence length.

From the Cooper-pair wave function $\Phi^{n}(R,r)$, one can extract the coherence length.
The coherence length gives information about the spatial extension of the pair.
In INM, this quantity can be simply approximated by the Pippard formula 
defined in Eq.~\ref{coher:pippard} within an acceptable level of accuracy. See Ref.~\cite{Matsuo_2006}
for a detailed discussion.
For a given density, the Cooper-pair extension in INM is uniform
and its value is inversely proportional to the pairing gap at the Fermi surface.
At variance, in finite nuclei, the neutron coherence length depends on the distance
$R$ from the center of the nucleus and it is defined as
\begin{equation}\label{coer-ws}
  \chi^{n}(R) = 
    \sqrt{\frac{\int d^{3}r \; r^{4} |\Phi^{n}(R,r)|^{2}}
               {\int d^{3}r \; r^{2} |\Phi^{n}(R,r)|^{2}}
         }.
\end{equation}
The neutron coherence length for $Z=40$ and $Z=50$ systems
is shown in Fig.~\ref{coherence_zrsn} for a wide range of the neutron number. 
In the case of finite nuclei (i.e., $\elemA{120}{Zr}$ and $\elemA{120}{Sn}$),
there is a minimum at the nuclear surface, as recent studies already discussed
\cite{Pastore_2008,Vinas_2010a,Pillet_2010,Pillet_2007}.
At the inner-crust densities this nuclear-surface effect gradually disappears as 
one moves from the outermost layers toward the star core
(i.e., from $\elemA{180}{Zr}$ and $\elemA{500}{Zr}$).
In the outer neutron gas (i.e., large $R$), WS-cell calculations recover
the INM coherence length value at the corresponding density, as it can be seen
from the arrows in Fig.~\ref{coherence_zrsn}. These arrows correspond to the Pippard coherence
length $\chi^n_P$ computed with the Fermi momentum $k_{F,n}$ of the outer gas of the given inner-crust region
(see Table~(\ref{tabWS})) and with $\Delta_n(k_{F,n})$ taken from the INM results in the left panel
of Fig.~\ref{pairing_gaps}.

At present, the relation between the coherence length and the strength of the pairing correlations
still needs to be better clarified.
According to recent studies in finite nuclei \cite{Pastore_2008,Vinas_2010a,Pillet_2010,Pillet_2007},
the coherence length has a very small dependence on the strength of the pairing interaction
and, consequently, on the value of the pairing gap at the Fermi level.
This lead to the conclusion \cite{Pillet_2010} that
the minimum of the coherence length in finite nuclei has little to do with
an enhanced strength of the pairing correlations at the nuclear surface.
The minimum is a finite-size effect.

A possible explanation is that the superfluid properties are mostly 
determined by the levels around the Fermi energy (typically in the region
$\varepsilon_F \pm \Delta$). 
At variance with the INM case, where a continuum of states
in the region $\varepsilon_F \pm \Delta$ contributes to the pairing correlations,
in finite nuclei close to the stability valley
the dependence of the coherence length on the pairing gap at the Fermi level
is washed out by the shell structure.
The situation is different for the dineutron halo nucleus $\elemA{11}{Li}$ \cite{Hagino_2010}.
The weakly-bound valence single-particle wave functions extend far outside the core and
couple with the continuum. As a consequence, finite-size effects are not able to suppress the dependence
of the coherence length on the pairing-interaction strength.
The coherence length in $\elemA{11}{Li}$ has a minimum at the nuclear surface,
but this minimum disappears with a negligible pairing interaction.

In the same way as $\elemA{11}{Li}$, we can then expect the coherence length
in the inner crust to depend on the pairing-interaction strength,
as the Fermi energy lies in the continuum.
Following the ideas of \cite{Hagino_2010, Duguet_private, Pillet_2010} in finite nuclei, 
we investigated this dependence in the $\elemA{180}{Zr}$ and $\elemA{1800}{Sn}$ WS cells,
whose neutron coherence length
is shown in Fig.~\ref{coherence_pairingstrength} for the SLy4 Skyrme functional and with the 
$\vlowk$ pairing interaction rescaled by a factor $x$. 
The curves in each panel of Fig.~\ref{coherence_pairingstrength} correspond to different
rescaling factors $x$. From top to bottom, pairing correlations increase, with consequently
larger and larger LCS pairing gaps.
The coherence length in the inner crust depends on the strength of the pairing interaction. 
This dependence is present also inside the proton cluster and it is
stronger in the outer neutron gas, where the inverse proportionality to the 
pairing gap is recovered (see Eq.~\ref{coher:pippard}).

We conclude that the coherence length can depend on the strength of the pairing interaction,
even inside the nucleus, but finite-size effects suppress this dependence when the Fermi level
does not lie close to the continuum.

\subsection{Pairing field}\label{Sect:results_pairing_field}

\begin{figure}
\begin{center}
\includegraphics[width=0.4\textwidth,angle=0,clip=]{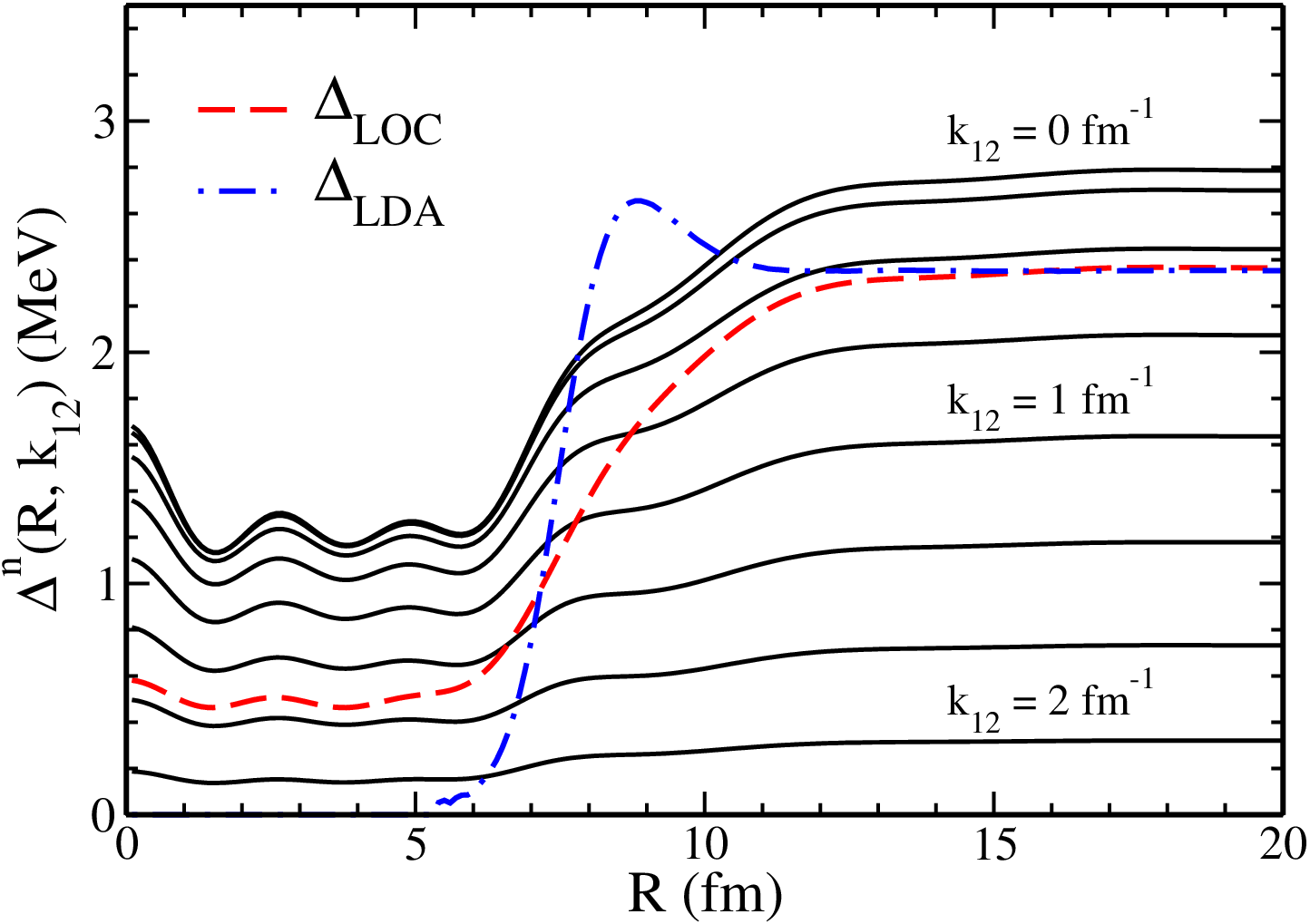}
\end{center}
\caption{(Color online) Neutron pairing field $\Delta^{n}(R,k_{12})$ 
for $\elemA{1350}{Sn}$ (solid lines) as a 
function of $R$ for fixed values of the relative momentum $k_{12}$.
The dashed line corresponds to the local neutron pairing field $\Delta^{n}_{\text{LOC}}(R)$ 
defined in Eq.~\ref{eq:local_pairing_field}. 
The dot-dashed line corresponds to the LDA approximation \cite{Margueron_2007}.} 
\label{fourier1350sn}
\end{figure}

\begin{figure*}
\includegraphics[width=0.40\textwidth,clip=]{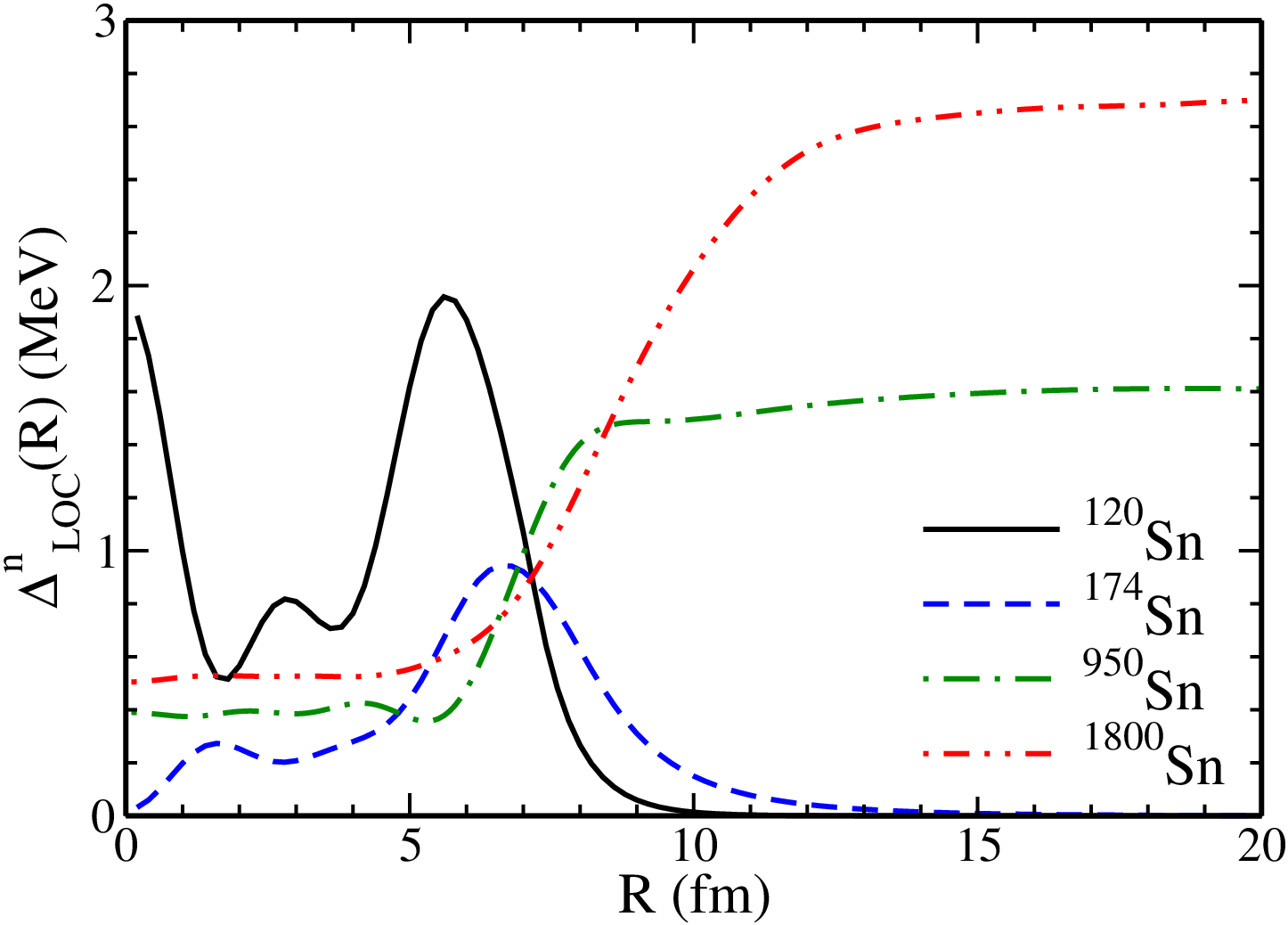}
\includegraphics[width=0.40\textwidth,clip=]{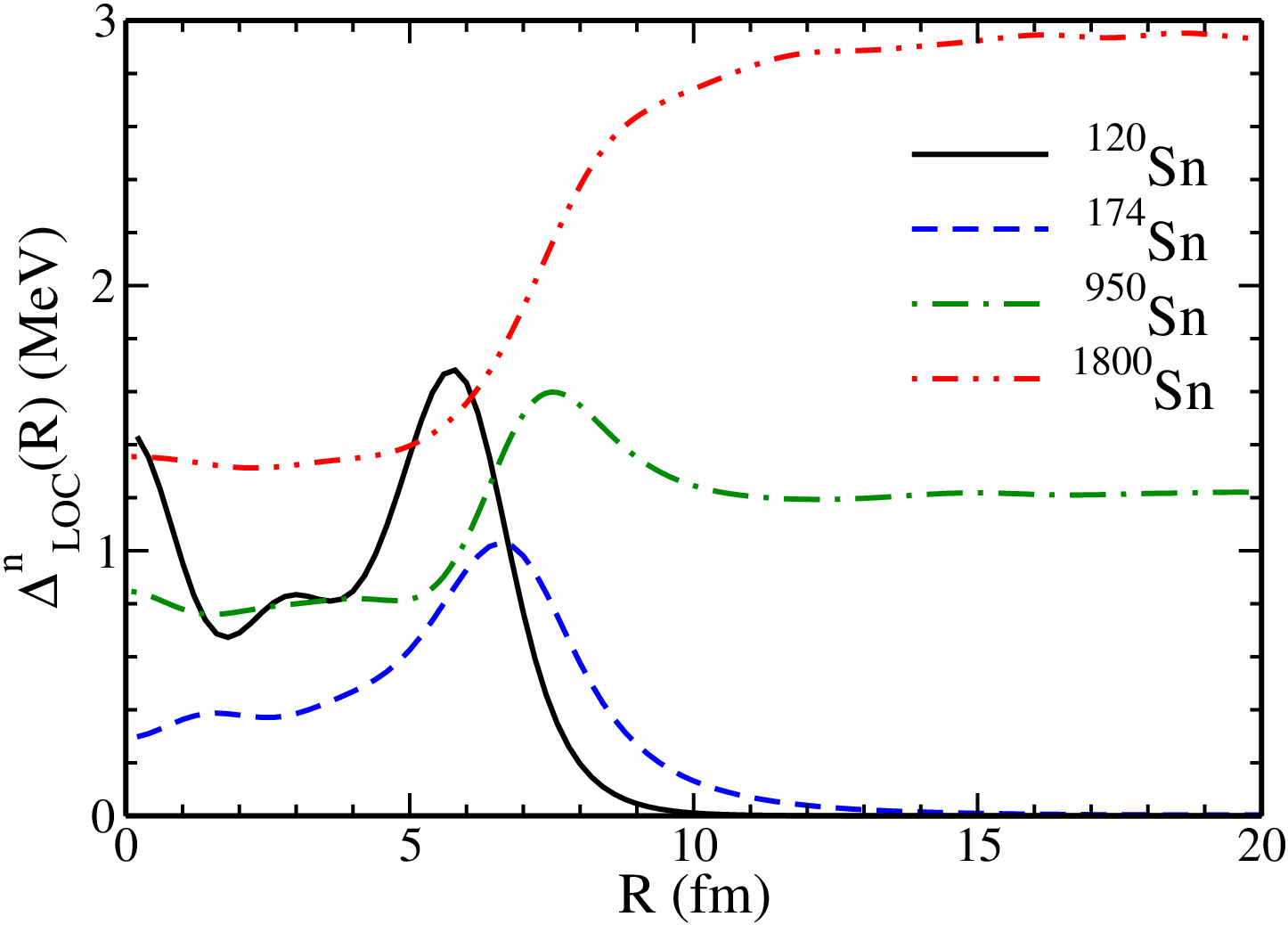}
\caption{(Color online) Neutron local pairing field for $Z=50$ systems, obtained with $\vlowk$
(left panel) and the DDDI (right panel) pairing interactions on top of the SLy4 Skyrme functional.}
\label{local_pairing_vlowk_DDDI}
\end{figure*}

In this section we investigate the spatial properties of the pairing field.
For a local pairing interaction $v(\mathbf{r}_{1}-\mathbf{r}_{2})$, the
pairing field reads \cite{Pastore_2008}

\begin{equation}
\Delta^{q}(\mathbf{r}_{1},\mathbf{r}_{2})=
  -v(\mathbf{r}_{1}-\mathbf{r}_{2})\Phi^{q}(\mathbf{r}_1,\mathbf{r}_2),
\end{equation}
where $\Phi^{q}(\mathbf{r}_1,\mathbf{r}_2)$ is the Cooper pair wave function defined 
in Eq.~\ref{a:density:r}. 
It is convenient to perform a Wigner transform \cite{Ring_1980}, changing to 
the center-of-mass coordinates and performing a Fourier transform on the relative distance $r$.
In this way we obtain the pairing field $\Delta^{q}(\mathbf{R},\mathbf{k}_{12})$, 
as a function of the two-particle center of mass $\mathbf{R}$ and of their relative momentum $\mathbf{k_{12}}$.
The pairing field depends weakly on the relative angle between the vectors 
$\mathbf{k}_{12}$ and $\mathbf{R}$, so we show the results after performing an angular average.

The neutron pairing field $\Delta^{n}(R,k_{12})$ is shown in Fig.~\ref{fourier1350sn} for the 
$\elemA{1350}{Sn}$ WS cell, using the SLy4 functional and the $\vlowk$ pairing interaction. 
We observe that the pairing field $\Delta^{n}(R,k_{12})$, has a strong dependence on 
the relative momentum $k_{12}$
and it is suppressed at the center of the cell by the presence of the nucleus. 
As a consequence one observes a global reduction of the pairing gap 
(cf. Sect.~\ref{Sect:results_pairing_gaps}). 
Our results on the pairing field agree with the more detailed discussion of Ref.~\cite{Barranco_1998}.

A local approximation $\Delta^q_{LOC}$ to the pairing field can be obtained by defining a local
Fermi momentum $k_{F,q}(R)$
\begin{equation}
\frac{\hbar^{2}k^{2}_{F,q}(R)}{2m_q^{*}(R)} \equiv \varepsilon_{F,q}-U^{q}_{centr}(R),
\end{equation}
and then taking the pairing field value
\begin{equation}\label{eq:local_pairing_field}
\Delta^{q}_{\text{LOC}}(R)\equiv \Delta^{q}(R,k_{F,q}(R)).
\end{equation}
$U^{q}_{centr}(R)$ is the central potential of the single-particle Hamiltonian.

The neutron local pairing field $\Delta^{n}_{\text{LOC}}$ is shown in Fig.~\ref{fourier1350sn},
together with an LDA approximation $\Delta_{LDA}$ to the pairing field \cite{Margueron_2007}.
$\Delta_{LDA}$ has been obtained 
solving the neutron gap equation (cf. Eq.~\ref{gapeqINM}) in asymmetric nuclear matter,
at different values of the Fermi momentum $k_{F,n}(R)=(3\pi^{2}\rho_{n}(R))^{1/3}$, where
$\rho_{n}(R)$ is the HFB density for the $\elemA{1350}{Sn}$ cell.
The LDA pairing field is more suppressed in the interior of the nucleus (where 
$k_{F,n}\approx 1.33$ fm$^{-1}$), and it is peaked at the nuclear surface. 
The local pairing field, instead, has a monotonic behavior. 
The reason of the difference between the two approximations has already been analyzed in a 
previous work \cite{Vinas_2011} and comes from the fact that
the classical LDA approximation is not able to capture the shell-structure effects of the system.

The evolution of the local neutron pairing field from finite nuclei to the inner crust is shown
in Fig.~\ref{local_pairing_vlowk_DDDI} for $Z=50$ systems. 
The analysis is done for both $\vlowk$ (left panel) and the DDDI (right panel) pairing interactions.
The latter is a local interaction and the pairing field does not depend on $\mathbf{k_{12}}$ by construction.
As we go from finite nuclei to the inner crust, the role
of the nuclear surface changes drastically. The local pairing field is peaked at 
the surface of finite nuclei, at variance with
nuclei immersed in a sea of superfluid neutrons.
The value of the pairing field inside the nucleus represents the main difference between the results
obtained with the two pairing interactions of Fig.~\ref{local_pairing_vlowk_DDDI}.

Although the pairing field and its local approximation can be interpreted as a measure
of the strength of the pairing correlations, one has to keep in mind that these are not
observables. Hence, the association of a small coherence 
length (i.e., close correlated nucleons, see Fig.~\ref{coherence_zrsn}) with a large pairing field
(i.e., large pairing correlations, see Fig.~\ref{local_pairing_vlowk_DDDI}) has to be taken
with a grain of salt. The suppression of the dependence of the coherence length on the strength
of the pairing interaction, due to shell-structure effects in finite nuclei
makes the above association at least unclear.

\section{Conclusions} \label{Sect:conclusions}

We studied the superfluid properties of the inner crust of neutron stars, representing 
the different density regions with 11 spherical WS cells. In the innermost layers close to the
star core, the WS approximation turned out to break down, even at the HF level, where
protons leak out of the center of the cell.
The method starts to be unstable at baryonic densities 
$\rho_b \approx 8 \cdot 10^{13}$ g $\cdot$ cm$^{-3}$ and it definitely breaks down at 
$\rho_b \approx 10^{14}$ g $\cdot$ cm$^{-3}$.
The WS approximation should not be used whereas these instabilities occur.

Within the limits of applicability of the method, we performed fully self-consistent HFB
calculations, based on Skyrme functionals plus realistic pairing interactions.
The pairing interaction has been taken as the low-momentum evolution of the
Argonne AV18 potential. We restricted our calculations to the
$\partialwave{1}{S}{0}$ pairing matrix elements, as higher partial waves are expected
to give a much smaller contribution in the inner crust.

From a comparison with INM, the presence of the protons in the inner crust is found to reduce 
the pairing gap at the Fermi surface of about $10 \%$ at its maximum. This suppression
is negligible below $k_{F,n}=0.5\fmi$. 
We also compared two different WS ($R_{WS},Z$) configurations, obtained from two 
energy-minimization procedures \cite{Negele_1973, Baldo_2005}, with the results
lying in a very narrow band.

The results obtained with effective pairing interactions
(namely, the Gogny D1 interactions and a density-dependent contact force)
differ substancially from the results obtained with the realistic pairing interactions.
Not only the non-arbitrariness of the latter ensures a higher reliability,
but also the availability of a high-precision separable representation of
low-momentum realistic potentials makes the calculations feasible even in the inner
crust of neutron stars.
Hence, we advocate for adopting these realistic pairing interactions to get more reliable
results when applying microscopic EDF theories to the inner crust of neutron stars. 

Higher-order pairing correlations are expected to play an important role
in the inner crust of neutron stars, where the exchange of collective
vibrations leads to a repulsive interaction and, consequently, to a suppression of the pairing gap
\cite{Gori_2004, Baroni_2008}. 
At sufficiently high densities ($k_{F,n} \ge 0.7 \fmi$), three-body forces are also
expected to contribute to the pairing interaction \cite{Hebeler_2010}.
A quantitative assessment of this effect requires all contributions
to be treated consistently and represents the subject of a future work.
Recently, ab-initio Quantum Monte Carlo calculations have been carried out for low-density
infinite neutron matter ($k_{F,n} \le 0.5 \fmi$) \cite{Gezerlis_2010},
with an effort to understand the differences with other Monte Carlo results 
\cite{Fabrocini_2005, Gandolfi_2008, Abe_2009}.
A quenching of the mean-field pairing gap is predicted, resulting in a larger pairing gap
than what previous calculations found
\cite{Chen_1993, Wambach_1993, Schulze_1996, Schwenk_2003, Cao_2006}.

In this work we also presented a detailed study of the Cooper-pair spatial properties, 
which showed how the coherence length in the inner
crust depends on the strength of the pairing interaction.
This dependence is present also inside the proton cluster and it is
stronger in the outer neutron gas, where the inverse proportionality to the 
pairing gap is recovered (see Eq.~\ref{coher:pippard}).
This result was expected at the inner-crust densities, where the Fermi energy lies
in the continuum, at variance with nuclei close to the
stability valley, where shell-structure effects suppress this dependence.
At present, the relation between the coherence length and the strength of the pairing
correlations still needs to be better clarified.

\section{Acknowledgments}

We are grateful to T. Lesinski and T. Duguet for useful discussions 
and for providing us with the separable representation of low-momentum realistic interactions.
This work was supported in part by the
Academy of Finland and the University of Jyv\"askyl\"a within the FIDIPRO program
and
the Natural Sciences and Engineering Research Council of Canada (NSERC). TRIUMF
receives funding via a contribution through the National Research Council Canada.


\end{document}